\documentclass[aap]{stylefile}

\usepackage{hyperref}
\usepackage{bbm, enumerate, mathtools, subcaption}
\usepackage{ulem}
\usepackage{dsfont}

\usepackage{cancel}
\usepackage{blindtext}

  \usepackage{pdflscape}

\usepackage{mathtools}
\usepackage{amsmath,amssymb}
\usepackage{hyperref}
\usepackage{amsthm}
\usepackage{arydshln}
\usepackage{enumitem}
\usepackage{xcolor}
\usepackage{threeparttable}
\usepackage{tikz}
\usetikzlibrary{shapes.arrows}

\usepackage{pgfplots}
\pgfplotsset{compat=1.15}
\usepackage{adjustbox}
\usepackage{bm}
\usepackage{gincltex}
\usepackage{caption}
\usepackage{subcaption}
\usepgfplotslibrary{fillbetween}
\usepgfplotslibrary{groupplots}
\usetikzlibrary{plotmarks}
\usetikzlibrary{patterns}
\pgfplotsset{
tick label style={font=\footnotesize},
label style={font=\footnotesize},
legend style={font=\footnotesize},
}

\newtheorem{theorem}{Theorem}[section]
\newtheorem{corollary}[theorem]{Corollary}

\newtheorem{remark}[theorem]{Remark}

\let\plainqed\qedsymbol
\newcommand{\claimqed}{$\lrcorner$}

\DeclareMathOperator*{\argmin}{arg\,min}

\usepackage[utf8]{inputenc}

\newcommand{\EE}{\mathbb E }
\newcommand{\ti}{(t) }

\newcommand{\q}{{\bf Q}}

\allowdisplaybreaks

\begin{document}

\begin{frontmatter}
\title{Heavy-traffic Optimality of Skip-the-Longest-Queues in Heterogeneous Service Systems}

\begin{aug}
\author{\fnms{Yishun} \snm{Luo}\ead[label=e2]{luo00329@umn.edu}}
\and
\author{\fnms{Martin} \snm{Zubeldia}\ead[label=e1]{zubeldia@umn.edu}}
\address{
University of Minnesota, Minneapolis, USA.}

\end{aug}

\begin{abstract}
We consider a discrete-time parallel service system consisting of $n$ heterogeneous single server queues with infinite capacity. Jobs arrive to the system as an i.i.d. process with rate proportional to $n$, and must be immediately dispatched in the time slot that they arrive. The dispatcher is assumed to be able to exchange messages with the servers to obtain their queue lengths and make dispatching decisions, introducing an undesirable communication overhead. In this setting, we propose a ultra-low communication overhead load balancing policy dubbed $k$-Skip-the-$d$-Longest-Queues ($k$-SLQ-$d$), where queue lengths are only observed every $k(n-d)$ time slots and, between observations, incoming jobs are sent to a queue that is not one of the $d$ longest ones at the time that the queues were last observed. For this policy, we establish conditions on $d$ for it to be throughput optimal and we show that, under that condition, it is asymptotically delay-optimal in heavy-traffic for arbitrarily low communication overheads (i.e., for arbitrarily large $k$).
\end{abstract}


\end{frontmatter}

\setcounter{tocdepth}{2}
\tableofcontents

\section{Introduction}

Parallel service systems are ubiquitous in many applications, from checkout lines at the supermarket, to server farms for cloud computing. These systems involve a single stream of incoming jobs that are dispatched to one of many single-server queue operating in parallel. In these systems, the main performance metrics of interest are: the maximum throughput that can be achieved, and the average delay of a typical job. In this setting, a throughput and delay optimal policy is the well-know Join-the-Shortest-Queue (JSQ) policy, where incoming jobs are sent to one of the shortest queues. Despite its outstanding performance, in practice this policy does not scale well to large systems, as the communication overhead required to implement the JSQ policy becomes prohibitive. In order to make dispatching policies scalable to large systems, many policies have been proposed in the past. In particular, it was discovered that the use of memory at the dispatcher can greatly decrease the communication overhead require to implement policies with excellent throughput and delay performance, with this communication overhead being the more expensive requirement in most applications. Since an exact analysis for these more sophisticated policies is usually intractable, it is common to consider the performance in several asymptotic regimes.

In this paper, we focus on the throughput and delay performance in classical and many-server heavy-traffic regimes, where the arrival rate converges to the maximum capacity of the system. In such regimes, the multi-dimensional state of the queuing system is reduced to a single  dimensional state. This behavior of the queuing system is known as the State Space Collapse (SSC) and it is critical to achieve delay optimality. In particular, we introduce and analyze a novel dispatching policy that attains throughput and delay optimality while having an arbitrarily small communication overhead. Before presenting our main contributions, we briefly outline prior work in this setting. 

There is a large body of literature on load balancing policies for parallel service systems. On one extreme, Round Robin has the lowest delay among all policies with no communication overhead \cite{roundRobin}, although its delay performance is far from optimal, and its stability region is reduced. On the other extreme, Join-the-Shortest-Queue (JSQ) policy is known to be delay optimal for homogeneous systems \cite{winston1977optimality-JSQcontinuous} and throughput optimal in general \cite{bramson11}, but it requires prohibitive amounts of communication overhead. Motivated by the need to reduce communication overhead, there are many policies that fall between Round Robin and JSQ, such as the Power-of-$d$-Choices (Po$d$) \cite{mitzenmacher,vvedenskaya,BorstPowerOfd}, Join-Idle-Queue (JIQ) \cite{badonnelBurgess,lu2011join-JIQ,stolyar14}, and other memory enhanced policies that make dispatching decisions based on stored information, such as server ID, queue length, etc. \cite{shah2002use-memeory-JSd,gamarnik2018delay-memory-1,gamarnik2022stability-memory-hetero,van2020zero-JIQ-open, vargaftik2020lsq-localshortestqueue}.  However, policies like Po$d$ and JIQ may exhibit poor performance in heterogeneous systems \cite{gardner2021scalable-PodJIQ-poor-JSQJIQheterogeneous}. To address this issue and improve performance, modified versions of JSQ \cite{gardner2021scalable-PodJIQ-poor-JSQJIQheterogeneous, bhambay2022asymptotic-speedawareJSQ,weng2020optimal-JFSQ-JISQ}, Po$d$ \cite{abdul2022general-Podgeneralpolicy} and JIQ \cite{gardner2021scalable-PodJIQ-poor-JSQJIQheterogeneous,weng2020optimal-JFSQ-JISQ} have been developed that incorporate service speed information into the dispatching decisions. Furthermore, if the servers themselves are capable of performing analysis, there exist policies, such as CARE \cite{mendelson2022load-estimationtriggered-general}, that make dispatching decisions based on estimated queue lengths.

Since most load balancing policies are not amenable to exact analysis, several asymptotic regimes have been considered in the literature. One of the most popular is classic heavy-traffic \cite{halfinWhitt}, where the load of the queueing system is made to converge to one. For this regime, JSQ has been shown to be throughput and delay optimal in classical and many-server heavy-traffic in continuous \cite{weber1978optimal-JSQcontinuous, winston1977optimality-JSQcontinuous} and discrete time setting \cite{eryilmaz2012asymptotically-driftmethod, hurtado2022heavy-JSQequaloptimality}. Moreover, the Po$d$ policy has also been shown to be delay optimal in heavy-traffic \cite{eryilmaz2012asymptotically-driftmethod,hurtado2020transform-Po2optimal,hurtado2021throughput-Pod-general,maguluri2014heavy-heavytraffic-po2jsqresourceallocation}. While the original JIQ policy is not delay-optimal in heavy-traffic \cite{zhou2017designing-JIQnotoptimal}, a modified version of it, dubbed JBT-$r$, has been shown to be optimal \cite{zhou2018heavy-JBT-r-general}. JBT-$r$ has also been shown numerically to achieve a low communication overhead, although it lacks a closed-form expression for it. Finally, it has been shown in \cite{zhou2021asymptotically-localestimated, zhou2019flexible-MultiDimensionSSC-generaldeltaPpoliocy, zhou2018degree-ideaoftheorder} that any load balancing policy is heavy-traffic delay optimal as long as it satisfies certain conditions. Nevertheless, these conditions are hard to verify in practice for many policies, especially for heterogeneous systems and for ultra-low communication overhead policies like the one we introduce in this paper. Another important regime is many-server heavy-traffic regime where both the number of servers and arrival rate increase together. JSQ and Po$d$ have been shown to be delay optimal in this regime under certain conditions \cite{hurtado2022load-manyserverheavytraffic,zhou2020note-manyserver}.

\subsection{Our contribution}

In this paper we focus on developing a family of dispatching policies that are throughput and delay optimal in heavy traffic, while requiring extremely low communication overhead. Compared with the policies mentioned above, our family of policies is not only easy to implement but also guarantee arbitrarily low communication overhead. Such policies are born from the following insights:
\begin{itemize}
    \item[(i)] In heavy traffic, queues are so long that they appear to be of an approximately constant length for long intervals of time. Therefore, there is no need to sample them in each time slot in order to make good dispatching decisions. 
    \item[(ii)] When making repeated dispatching decisions over many consecutive time slots using the same queue length vector, policies that work well over one time slot (like Joint-the-Shortest-Queue) do not work so well. For repeated decisions, it is better to spread out the arrivals among different queues as much as possible in order to maintain the balancedness of queues. 
\end{itemize}
With these insights in mind, we introduce a new load balancing policy called $k$-Round-Robin-Skip-the-Longest-$d$-Queues ($k$-SLQ-$d$), where $k$ and $d$ are positive integers. Under this policy, the queue lengths are sampled every $k(n-d)$ time slots, and incoming jobs between the sampling times are sent in a Round-Robin fashion to the $n-d$ queues that were shortest at the last time that they were sampled. This Round-Robin loop is done for $k$ rounds, which fills the $k(n-d)$ time slots in between queue length samplings. For this policy:
\begin{enumerate}
    \item We obtain its stability region and establish necessary and sufficient conditions on the number of skipped servers $d$ for throughput optimalilty when the servers are heterogeneous. In particular, we show that these do not depend on the number of rounds $k$.
    \item We show that, under almost the same condition on $d$ as for throughput optimality, the $k$-SLQ-$d$ policy is delay-optimal in classical heavy-traffic for any $k$. In particular, this implies that we can obtain delay optimality with an arbitrarily low communication overhead. Moreover, if the message rate can be made to depend on the stability slack $\epsilon$, we show that delay optimality can be achieved with a vanishing message rate of order $o(\epsilon)$.
    \item We extend our results to the many-server heavy-traffic regime. We show that the number of skipped servers $d$ in order to be throughput optimal is always of order $\Theta(n)$ unless the system is almost homogeneous. Moreover, we show that our policy is delay optimal in many-server heavy-traffic as long as $\epsilon \in o(n^{-11})$.
    \item For the homogeneous case, there is a phase transition in the delay performance in both classical and many-server heavy-traffics between using a pure Round-Robin policy, and doing Round-Robin skipping just the longest queue at the beginning of each round.
\end{enumerate}
From the methodological side, we use Lyapunov drift arguments to establish our results. However, the nature of our policy makes it so the one-step drift of the queue lengths can be positive at times. To overcome this challenge, we look at the drift of the queues over a whole cycle of $k(n-d)$ time slots. This brings new challenges, as usual tricks in the literature that are used to avoid problematic terms do not work over multiple time slots.

\subsection{Basic Notations}

We use $\mathbb R$ to denote the set of real numbers and $\mathbb R_+$ to denote the set of non-negative real numbers. Also, $\mathbb N$ denotes the set of natural numbers. Similarly, $\mathbb R^d$ denotes the set of $d$-dimensional real vectors. We use bold letters to denote vectors and, for any vector ${\bf x}$, we use $x_i$ to denote the $i^{th}$ coordinate of ${\bf x}$. The inner product of two vectors ${\bf x}$ and ${\bf y}$ in $\mathbb R^d$ is defined as $\langle {\bf x},{\bf y}\rangle = {\bf x}^T {\bf y} = \sum_{i=1}^d x_iy_i$. For any vector $x\in \mathbb R^d$, the $\ell_2$-norm is denoted by $\|{\bf x}\|_2 = \sqrt{\langle {\bf x} , {\bf x}\rangle }$, the $\ell_1$-norm is denoted by $\|{\bf x}\|_1 = \sum_{i=1}^d |x_i|$, and the sum of its components is denoted by ${\bf x}_\Sigma = \sum_{i=1}^d x_i$. For any positive natural number $d$, $\bf{1}_d$ and $\bf{0}_d$ denotes the vector of all ones and vector of all zeros of size $d$ respectively. For ease of notation, at most places, we drop the subscript and just use $\bf{1}$ and $\bf{0}$ instead of $\bf{1}_d$ and $\bf{0}_d$.  For any set $A$, $\mathds{1}_A$ denotes the indicator random variable for set $A$. We use $\lfloor \cdot \rfloor$ and $\lceil \cdot \rceil$ to denote the floor and ceiling functions respectively. We denote $[n]=\{1,2,\dots, n\}$, and $\mathcal{P}_d([n])$ as the set of all subsets of size $d$ of $[n]$.

\section{Queueing model and new dispatching policy}
In this section, we will first introduce the discrete time queuing model, our $k$-Round-Robin Skip the Longest Queue policy ($k$-SLQ-$d$), and present our main results. 

\subsection{Discrete-time queueing model} \label{subsec:DTQS}

We consider a discrete time queueing system consisting of $n$ single-server FIFO queues of infinite capacity. We denote the queue length vector at the beginning of time slot $t$ by $\q(t)$, where $Q_i(t)$ is the length of the $i$-th queue. Jobs arrive to the system as an i.i.d. process $\{A(t)\}_{t\geq 0}$, with $\EE [A(1)] = n\lambda$, Var$(A(1)) = n\sigma_{\lambda}^2$, and $A(1)\leq n A_{\max}$ almost surely. When $A(t)$ jobs arrive, they are all dispatched to a queue with index $I^*(t)$ according to our $k$-SLQ-$d$ policy, which we will describe later. Once the jobs join the queue, the servers process up to ${\bf S}(t)$ jobs waiting in the queues, with $\EE[{\bf S}(1)] = \boldsymbol{\mu}$, Var$(S_i(1)) = \sigma_{\mu_i}^2$, and $S_i(1)\leq S_{\max}$ almost surely, for any $i \geq 1$. Moreover, we assume that there exists $\mu_{\max} \geq \mu_{\min} > 0$ and $\sigma^2_{\mu_{\max}} \geq \sigma^2_{\mu_{\min}} \geq 0$ such that $\mu_{\min} \leq \mu_i \leq \mu_{\max}$ and $\sigma^2_{\mu_{\min}} \leq \sigma^2_{\mu_i} \leq \sigma^2_{\mu_{\max}}$ for all $i\geq 1$. Similar to the arrival process, the potential services are also independent and identically distributed across time slots, and they are independent of the queue length vector. 

To describe the queue dynamics, we denote the ``dispatching action'' chosen by the dispatcher by ${\bf Y}(t) \in \{0,1\}^n$ such that, if incoming jobs are sent to queue $i$ at time $t$, we have $Y_i(t) = 1$ and $Y_{j}(t) = 0$ for all $j\neq i$. Using this, the queue length process is given by
\begin{equation}
\label{eq: jsq_lindley}
    \q(t+1) = [\q(t) +A(t){\bf Y}(t) -{\bf S}(t) ]^+ = \q(t) +A(t){\bf Y}(t) -{\bf S}(t) + {\bf U}(t) ,
\end{equation}
where the operation $[\,\cdot\,]^+$ above is used because the queue lengths cannot be negative, and the term ${\bf U}(t)$ are the unused services that arise because there might not be enough jobs to serve. Note that, the unused service term $U_i\ti $ is positive only if $Q_i(t+1) =0$, which implies $Q_i(t+1)U_i(t) =0$ for all $i$, or simply $\langle \q(t+1),{\bf U}(t) \rangle =0$ for any $t>0$. Also, the unused service cannot be larger than the service itself, so we have $0\leq U_i(t)\leq S_i(t) \leq S_{\max}$. We drop the dependence on $t$ to denote the variables in steady state, i.e., $\q$ follows the steady state distribution of the queue length process $\{\q(t)\}_{t=0}^\infty$.

\subsection{$k$-Round-Robin Skip-the-Longest-$d$ Policy ($k$-SLQ-$d$)} \label{subsec:k-SLQ-d}

We now present our $k$-SLQ-$d$ policy, for some integers $k\geq 1$, and $d\in\{1,2,\dots,n-1\}$. The operation logic of the policy policy is as follows:
\begin{enumerate}
    \item In time slots $t$ that are multiple of $k(n-d)$, the dispatcher collects the queue length vector $\mathbf{Q}(t)$ and finds the IDs of the longest $d$ queues, breaking ties uniformly at random. Let $I_d(t)$ be the set of such indices.
    \item During the next $k(n-d)$ time slots (including time slot $t$), arriving jobs are sent to queues in a Round-Robin fashion over the set of indices $\{1,2,\dots,n\}\backslash I_d(t)$, which has cardinality $n-d$.
\end{enumerate}
To understand this policy, let us first consider the case of $k=1$. Here, choosing $d=n-1$ corresponds to the delay-optimal JSQ policy, and choosing $d=0$ would correspond to a pure Round-Robin policy\footnote{In this case the queue length vector would be queried every $n$ time slots, although its information would not be used.}. For intermediate values of $d$, incoming jobs during each consecutive $n-d$ time slots are sent to the $n-d$ queues that were shortest at the time that the queues were last queried, and the queue length vector is only queried once every $n-d$ time slots. Therefore, different values of $d$ give us policies that are somewhat in between JSQ and RR, each one striking a different balance between the communication overhead, and the goodness of the allocation of jobs.

We model this policy as a discrete-time Markov chain  $\{\mathbf{X}(t)\}_{t\geq 0}$ such that
\begin{align}
    \mathbf{X}(t) := \Big(\mathbf{Q}(t), I_d(t), I_R(t), N_R(t)\Big),  \label{def:Mkchain}
\end{align}
where $\mathbf{Q}(t) \in \mathbb{Z}_+^n$ is the queue length vector at time $t$, $I_d(t)\in\mathcal{P}_d([n])$ is the index set of the longest $d$ queues at time $t-[t \mod k(n-d)]$, $I_R(t)\in [n]$ is the index of the queue to which arriving jobs at time slot $t$ are sent, and $N_R(t) \in \{0,1,\dots,k-1\}$ is the number of Round-Robin rounds that have been done since time $t-[t \mod k(n-d)]$. Moreover, we also consider the discrete-time Markov chain $\big\{\tilde{\mathbf{Q}}(t)\big\}_{t\geq 0}$ defined as
\[ \tilde{\mathbf{Q}}(t) :=\mathbf{Q}(k(n-d)t), \]
which is the sampling of the original queue length process at the times that the queues are observed.

\subsubsection{Communication overhead} \label{subsec:frequency}

Under the $k$-SLQ-$d$ policy, the queue lengths are only sampled once every $k(n-d)$ time slots. Each time they are sampled, $2n$ messages are exchanged. Therefore, the average number of messages per job is
\begin{align}
    \text{Message rate} = \frac{2}{\lambda k(n-d)}. \notag 
\end{align}
The idea is that increasing the number of rounds $k$ can dramatically decrease the communication overhead of the algorithm. As we will see, increasing $k$ has no effect on the stability region of the algorithm, and the delay-performance in heavy-traffic is optimal for any finite $k$. Moreover, choosing a smaller $d$ can also dramatically decrease the communication overhead by making each round longer. However, we will show that this will have an impact on the stability region.

\section{Main results}

\subsection{Stability region} \label{subsec:stabreg}

The parallel queueing system that we consider is stabilizable as long as the arrival rate $\lambda$ is in the set
\begin{align} \label{def:capacityregion}
    C := \left\{ \lambda \in \mathbb{R}^+: n \lambda < \sum\limits_{l=1}^n \mu_l \right\}
\end{align}
In fact, the the $1$-SLQ-$(n-1)$ policy (which is equivalent to the JSQ policy) has this capacity region. The following theorem shows that the stability region of a general $k$-SLQ-$d$ policy does not depend on $k$, and it is increasing in $d$.

\begin{theorem}\label{thm_our:stable}
    Fix $k \geq 1$ and $d\in \{1,\dots,n-1 \}$. Under the $k$-SLQ-$d$ policy, the Markov Chain $ \{\mathbf{X}(t)\}_{t\geq 0}$ is positive recurrent if and only if:
    \begin{align}
        n\lambda < \min_{I \subseteq [n] \, : \, |I| \geq d+1 } \left\{  \frac{n-d}{|I|-d} \sum_{ l\in I} \mu_l \right\}. \label{eq:feasibleregion}
    \end{align}
\end{theorem}

\noindent The proof relies on a multi-step version of the Foster-Lyapunov theorem \cite{dai2020processing-statedependentMC} to prove positive recurrence when the arrival rate is small enough. However, while most multi-step Lyapunov drift arguments are utilized to overcome some averaging effect in the drift (such as in \cite{zubeldia2025matching}), we need to use a multi-step argument because our policy is implemented across multiple time steps. Moreover, the proof relies on a stochastic comparison with a simpler process to show null-recurrence or transience when the arrival rate is large enough. The complete proof is given in Section~\ref{prf:thm_stable}.\\

The following corollary provides a condition under which the stability region is maximal.

\begin{corollary}\label{co_our:condition}
    The Markov chain $ \{\mathbf{X}(t)\}_{t\geq0} $ is positive recurrent for all $ n\lambda < \sum_{l=1}^n \mu_l $ if and only if 
    \begin{align}
        \sum\limits_{l=1}^{n} \mu_l \geq (n-d) \max\limits_{l\in[n]} \big\{ \mu_l \big\}. \notag
    \end{align}
\end{corollary}
The proof is given in Section~\ref{prf:cor_stableregion}.\\

Note that, for any rate vector, we can always pick a $d$ large enough such that the system is stable for all arrival rates $ n\lambda < \sum_{l=1}^n \mu_l $. Indeed, this is true if and only if
\[ d \geq n - \frac{1}{\max\limits_{l\in[n]} \big\{ \mu_l \big\}} \sum\limits_{l=1}^n \mu_l. \]
In particular, in order for skipping just the longest queue (i.e., $d=1$) to be enough for the system to be throughput optimal, we need 
\[  \frac{1}{n - 1} \sum\limits_{l=1}^n \mu_l \geq  \max\limits_{l\in[n]} \big\{ \mu_l \big\}. \]
That is, we need the system to be close to homogeneous. Moreover, note that the stability of the system does not depend on the number of rounds $k$.\\

For the rest of this section, we assume that
\begin{align}\label{eq:d_large_enough}
d > n - \sum\limits_{l=1}^n \frac{ \mu_l}{\max\limits_{l\in[n]} \big\{ \mu_l \big\}},
\end{align}
so that the system under $k$-SLQ-$d$ is stable for all $n\lambda < \sum_{l=1}^n \mu_l$. Under this assumption, we define the capacity slack as
\begin{align}
    \epsilon := \sum\limits_{l=1}^n \mu_l - n\lambda.
\end{align}
In what follows, we focus on the asymptotic behavior of the steady-state queue lengths in the limit as $\epsilon$ converges to zero, i.e., in the heavy-traffic regime. In particular, we will show that $k$-SLQ-$d$ is asymptotically delay-optimal in heavy-traffic.

\subsection{State Space Collapse} \label{subsec:SSC}

In this subsection, we show that the queue length vector $\mathbf{Q}$ (i.e., the steady state queue length vector at the time that the queues are sampled) satisfies a form of State Space Collapse in heavy-traffic. That is, the $n$-dimensional steady-state queue length vector collapses along the $1$-dimensional subspace defined by the vector $\mathbf{c} := ( 1 / \sqrt{n} )_{l=1}^{n}$, in heavy-traffic. We define the projection of the queue length vector along this vector as
\begin{align}
    \mathbf{Q}_{\parallel}(t) = \left < c,\mathbf{Q}(t) \right > c=\left ( \frac{\|\mathbf{Q}(t)\|_1}{n} \right )_{l=1}^n. \notag
\end{align}
Moreover, we define the perpendicular component as
\begin{align}
     \mathbf{Q}_{\perp}(t) = \mathbf{Q}(t) - \mathbf{Q}_{\parallel}(t) = \left ( \tilde{Q}_l(t) - \frac{\|\mathbf{Q}(t)\|_1}{n} \right )_{l=1}^n. \notag
\end{align}
In the theorem below, we show that the second moment of the steady-state random variable $\|\mathbf{Q}_{\perp}\|_2$ is upper bounded by a constant.

\begin{theorem}\label{thm_our:SSC}
    Consider the $k$-SLQ-$d$ system as defined in Section \ref{subsec:k-SLQ-d}. Suppose that Equation \eqref{eq:d_large_enough} holds, and that
    \[ \epsilon = \sum\limits_{l=1}^n \mu_l - n\lambda \leq \frac{\min \left \{ \mu_{\min}, \left ( \frac{1}{n-d} \sum\limits_{i=1}^n \mu_i \right ) - \mu_{\max} \right \}}{2}. \]
    Then, there exists a function $N_2(n,k,d)$, independent from the slack $\epsilon$, such that
    \[ E\left[ \left\| \mathbf{Q}_{\perp} \right\|_2^2 \right ] \leq N_2(n,k,d), \]
    for all $n, k, d \geq 1$, with
    \begin{align}
        N_2(n,k,d) \in \Theta\big( n^8 (n-d)^2 k^2 \big). \label{eq:cor_N_2}
    \end{align} 
\end{theorem}
The proof is given in Appendix \ref{prf:thm_SSC}.\\

On the other hand note that, for any dispatching policy, the sum of the queue lengths is lower bounded by the queue length of a single server queue with the same arrivals $A(t)$, and services $S(t)=\sum_{l+1}^n S_l(t)$. This implies that $\mathbb{E}[\|\mathbf{Q}\|_1] \in \Omega(1/\epsilon)$. Combining this with Theorem \ref{thm_our:SSC}, we get that
\[ \lim_{\epsilon\to 0} \frac{\mathbb{E}[\|\mathbf{Q}_{\perp}\|_1]}{\mathbb{E}[\|\mathbf{Q}\|_1]} \leq \lim_{\epsilon\to 0} \frac{\sqrt{n}\mathbb{E}[\|\mathbf{Q}_{\perp}\|_2]}{\mathbb{E}[\|\mathbf{Q}\|_1]} \leq \lim_{\epsilon\to 0} \frac{\sqrt{n\mathbb{E}[\|\mathbf{Q}_{\perp}\|^2_2]}}{\mathbb{E}[\|\mathbf{Q}\|_1]} = 0. \]
This means that, when $\epsilon$ approaches zero, the perpendicular component of the queue length vector $\mathbf{Q}_\perp$ is negligible with respect to the queue length vector itself. Therefore, we have $\mathbf{Q} \approx \mathbf{Q}_\parallel$ when $\epsilon$ approaches zero, which can be interpreted as the collapse of the state space towards its projection onto a one-dimensional vector.

\begin{remark}
Now suppose that we make the number of rounds $k$ to be a function of the slack $\epsilon$. Since $N_2\in O(k^2)$ as a function of $k$, if $k\in o(1/\epsilon)$, then the State Space Collapse still holds. This means that, as long as the number of rounds is order-wise smaller than the queue lengths, then there will still be SSC. Indeed, as long as $k$ is smaller than the queue lengths, the imbalance created by these $k$ rounds will be negligible compared to the size of the queues.
\end{remark}

\subsection{Delay optimality in classical heavy-traffic} \label{subsec:upperbd}

In this subsection we exploit the SSC result in the previous subsection to establish the asymptotic delay optimality of $k$-SLQ-$d$. First, we provide an upper bound for the expected average queue length in steady state.

\begin{theorem} \label{thm_our:SSCUpper}
    Consider the $k$-SLQ-$d$ system as defined in Section \ref{subsec:k-SLQ-d}. Suppose that Equation \eqref{eq:d_large_enough} holds, and that
    \[ \epsilon = \sum\limits_{l=1}^n \mu_l - n\lambda \leq \frac{\min \left \{ \mu_{\min}, \left ( \frac{1}{n-d} \sum\limits_{i=1}^n \mu_i \right ) - \mu_{\max} \right \}}{2}. \]
    Then,
    \begin{align} \label{eq:thm_upper} 
        \epsilon E \left[ \frac{1}{n} \sum\limits_{l=1}^n Q_l \right ] \leq \frac{n \sigma_{\lambda}^2 + \sum\limits_{l=1}^{n}\sigma_{\mu_l}^2 }{2 n} + \epsilon^2 \frac{k(n-d)}{2n} + \epsilon k(n-d) \frac{2n A_{\max}+S_{\max}}{2} + \sqrt{\epsilon} \sqrt{N_2(n,k,d) n S_{\max}},
    \end{align}
    where $ N_2(n,k,d) $ is the one given in Theorem \ref{thm_our:SSC}.
\end{theorem}
The proof relies on a multi-step Lyapunov drift argument, and is given in Appendix \ref{prf:thm_upperbound}.\\

On the other hand, Lemma 5 in \cite{eryilmaz2012asymptotically-driftmethod} implies that, under any load balancing policy, we have
\[  \epsilon E \left[ \frac{1}{n} \sum\limits_{l=1}^n Q_l \right ] \geq \frac{n \sigma_{\lambda}^2 + \sum\limits_{l=1}^{n}\sigma_{\mu_l}^2 + \epsilon^2 - S_{\max} \epsilon }{2 n}. \]
Combining this with Theorem \ref{thm_our:SSCUpper} and Little's law, we get the following corollary.

\begin{corollary}
    Consider the $k$-SLQ-$d$ system as defined in Section \ref{subsec:k-SLQ-d}. Suppose that Equation \eqref{eq:d_large_enough} holds. Then, $k$-SLQ-$d$ is asymptotically delay optimal in heavy-traffic.
\end{corollary}

This corollary implies that, for any number of rounds $k$, the $k$-SLQ-$d$ policy is asymptotically delay-opptimal in heavy-traffic. However, we can go one step further and make $k$ a function of the slack parameter $\epsilon$, to have the same asymptotic delay performance with truly vanishing communication overhead.

\begin{corollary}
    Consider the $k$-SLQ-$d$ system as defined in Section \ref{subsec:k-SLQ-d}. Suppose that Equation \eqref{eq:d_large_enough} holds. Then, if $k\in o(1/\sqrt{\epsilon})$, then $k$-SLQ-$d$ is asymptotically delay optimal in heavy-traffic.
\end{corollary}

This implies that our policy can be asymptotically delay-optimal with a vanishing message rate of order $\omega\left(\sqrt{\epsilon}\right)$ messages per job. That is, the message rate converges to zero but at a rate that is strictly slower than $\sqrt{\epsilon}$.

\subsection{Join-the-Shortest vs Skip-the-Largest queue for homogeneous systems} \label{subsec:homogen}

In this subsection, we highlight the importance of skipping the longest queue compared to just having jobs join the shortest queue for a number of consecutive time slots. To showcase this, we consider a homogeneous system, that is, a system such that $\mu_l=\mu$ for all $l\geq 0$. In this case, skipping just the longest queue, i.e., setting $d=1$, is already enough for the system to be throughput optimal. We compare two extreme versions of our policy:
\begin{enumerate}
    \item The $k$-SLQ-$1$ policy, which does $k$ rounds of Round-Robin while skipping the queue that was the longest at the last time that the queues were sampled.
    \item The $k(n-1)$-SLQ-$(n-1)$ policy, which sends the next $k(n-1)$ arrivals to the queue that was shortest at the last time that the queues were sampled.
\end{enumerate}
Note that both of these policies have the same communication overhead of $2n/k(n-1)$ messages per time slot. However, their behavior is very different. In particular:
\begin{enumerate}
    \item $k$-SLQ-$1$ policy: The dispatcher sends $\Theta(kn)$ jobs to each of the $n-1$ shortest queues across $k(n-1)$ time slots.
    \item $k(n-1)$-SLQ-$(n-1)$ policy: The dispatcher sends $\Theta(kn^2)$ jobs to the same shortest queue across $k(n-1)$ time slots.
\end{enumerate}
Despite the fact that these two policies are both asymptotically delay-optimal, their pre-limit performances can be quite different when the number of servers $n$ is large. In order to see this, we perform numerical simulations for $n=\{10,20,50\}$,  $\mu = 2$, $\sigma_{\mu}=1$, $n\lambda= \{19.99, 39.98, 99.95 \}$, $n\sigma_{\lambda} = 5$. In table \ref{tab:sim_results_JSQ} we can see that the average queue length under repeated JSQ can be much larger than under SLQ. Moreover, we can see how going from a pure Round-Robin policy to a policy that skips the largest queue in each round can have a tremendous performance impact.
\begin{table}[htbp]
    \centering
    \begin{tabular}{|l|r|r|r|r|}
        \hline
        Policy & \# of servers  & Average queue & Stdev of queues & Message rate \\
        \hline \hline
        Round-Robin & 10 & 1738.597 & 237.0637 & 0.000 \\
        1-SLQ-1 & 10 & 231.805 & 5.051 & 1.111 \\
        9-SLQ-9 & 10 & 300.665 & 0.051 & 1.111 \\
        1-SLQ-1 & 20 & 109.541 & 10.643 & 1.053 \\
        19-SLQ-19 & 20 & 471.132 & 0.633 & 1.053 \\
        1-SLQ-1 & 50 & 144.778 & 26.550 & 1.020 \\
        49-SLQ-49 & 50 & 2537.674 & 2.904 & 1.020 \\
        \hline
    \end{tabular}
    \caption{Simulation results}
    \label{tab:sim_results_JSQ}
\end{table}

\section{Many-server heavy-traffic}

In this section we extend our results to the case where the number of servers $n$ diverges at the same time that the slack $\epsilon$ converges to zero. This is the so-called many-server heavy-traffic regime. Throughout this section, we assume that
\begin{align}
    \lim_{n\to\infty} \frac{1}{n} \sum\limits_{l=1}^{n}\mu_l = \bar{\mu} \qquad \text{and} \qquad \lim_{n\to\infty}  \frac{1}{n} \sum\limits_{l=1}^{n}\sigma_{\mu_l}^2 = \bar{\sigma}^2.
\end{align}

\subsection{Throughput optimality}
Recall that Corollary \ref{co_our:condition} implies that $k$-SLQ-$d$ is throughput optimal if and only if
\[ d \geq n - \sum\limits_{l=1}^n \frac{ \mu_l}{\max\limits_{l\in[n]} \big\{ \mu_l \big\}}. \]
Therefore, if
\[ \bar{\mu} < \lim\limits_{n\to\infty} \max\limits_{l\in [n]} \{\mu_l\} = \mu_{\max}, \]
then skipping the $d\in \Theta(n)$ longest queues is required to be throughput optimal. In particular, we need
\[ \lim_{n\to\infty} \frac{d}{n} \geq 1- \frac{\bar{\mu}}{\mu_{\max}} .  \]
This means that, in order to be throughput optimal, we need to skip a constant fraction of servers that only deppends on the ratio of the average service rate over the maximum one.

\subsection{Delay optimality} \label{subsec:manyserver}
The delay-optimality in the many-server heavy-traffic regime is obtained as a corollary of the non-asymptotic bounds of Theorem \ref{thm_our:SSCUpper}.

\begin{corollary} \label{co_our:manyserver}
    Consider the $k$-SLQ-$d$ system as defined in Section \ref{subsec:k-SLQ-d}. Suppose that Equation \eqref{eq:d_large_enough} holds. Then, if $\epsilon k^2 \in o\left(n^{-11}\right)$, then $k$-SLQ-$d$ is asymptotically delay optimal in heavy-traffic.
\end{corollary}

Note that we have three parameters here: the stability slack $\epsilon$ (which converges to zero), the number of servers $n$ (that diverges), and the number of rounds $k$ (which can either be a constant, or diverge). These need to be such that $\epsilon k^2 n^{11}$ converge to zero. Even when $k$ is constant, the system needs to be in a very heavy traffic with $\epsilon \in o\left(n^{-11}\right)$, well beyond Halfin-Whitt and Non-Degenerate-Slowdown, in order for the delay optimality to hold. In this case, the message rate is $\Theta\left(1/n\right)$, which is $o\left(\epsilon^{1/11}\right)$.

\section{Proof of Theorem \ref{thm_our:stable}} \label{prf:thm_stable}
First we prove the sufficient condition for positive recurrence. Suppose that
\begin{align}
    n\lambda < \min_{I \subseteq [n] \, : \, |I| \geq d+1 } \  \left \{ \frac{n-d}{|I|-d} \sum_{ l\in I} \mu_l \right \}. \notag 
\end{align}
We want to show the Markov Chain \hyperref[def:Mkchain]{$ \{\mathbf{X}(t)\}_{t\geq 0} $} is positive recurrent. Given that the dispatcher collects the queue length vector every $k(n-d)$ time slots under our $k$-SLQ-$d$ policy, we define a new Markov chain 
\[ \tilde{\mathbf{X}}(t) := \Big( \tilde{\mathbf{Q}}(t), \tilde I_d(t), \tilde I_R(t), \tilde N_R(t)\Big) \]
such that
\[ \tilde{\mathbf{X}}(t)=\mathbf{X}(k(n-d)t) \]
for all $t\geq 0$. Since the original Markov chain $ \{\mathbf{X}(t)\}_{t\geq 0} $ is irreducible, it is enough to prove that $\{\tilde{\mathbf{X}}(t)\}_{t\geq 0}$ is positive recurrent. The proof follows the same rough structure as in \cite{eryilmaz2012asymptotically-driftmethod}, but with considerably more technical challenges due to the nature of our policy.\\

Let us define the Lyapunov functions
\[ V(\mathbf{x})= \|\mathbf{x}\|_{2} \qquad \text{ and } \qquad  W(\mathbf{x})= \|\mathbf{x}\|_2^2, \]
and the drift operator $\Delta$ by
\begin{align}
    \Delta V(\mathbf{Q}(t)) & := V(\mathbf{Q}(t+1))-V(\mathbf{Q}(t)) \notag \\
    \Delta W(\mathbf{Q}(t)) & := W(\mathbf{Q}(t+1))-W(\mathbf{Q}(t)) \notag 
\end{align}
Further, the drift of $V$ can be bounded by the drift of $W$ as follows:
\begin{align}
    \Delta V(\mathbf{Q}(t)) & = \| \mathbf{Q}(t+1) \|_2 - \| \mathbf{Q}(t) \|_2 \notag \\
    & = \sqrt{\| \mathbf{Q}(t+1) \|_2^2} - \sqrt{\| \mathbf{Q}(t) \|_2^2} \notag \\
    & \leq \frac{1}{2\| \mathbf{Q}(t) \|_2} \Big( \| \mathbf{Q}(t+1) \|_2^2 - \| \mathbf{Q}(t) \|_2^2 \Big) \notag \\
    & = \frac{1}{2\| \mathbf{Q}(t) \|_2}\big[ \Delta W\big(\mathbf{Q}(t) \big) \big] \notag
\end{align}
Fix $t\geq 0$ such that $t$ is a multiple of $k(n-d)$, and let $t'=t/k(n-d)$.\\

We begin by showing that the drift of $V(\tilde{\mathbf{Q}}(t'))$ is uniformly absolutely bounded:
\begin{align}
    | \Delta V(\tilde{\mathbf{Q}}(t')) | & = |\| \tilde{\mathbf{Q}}(t'+1) \|_2 - \| \tilde{\mathbf{Q}}(t')\|_2 | \notag \\
    & \overset{(a)}{\leq} \| \tilde{\mathbf{Q}}(t'+1)-\tilde{\mathbf{Q}}(t') \|_2 \notag \\
    & \overset{(b)}{\leq} \| \tilde{\mathbf{Q}}(t'+1)-\tilde{\mathbf{Q}}(t') \|_1 \notag \\
    & = \| \mathbf{Q}(t+k(n-d))-\mathbf{Q}(t) \|_1 \notag \\
    & = \sum\limits_{l=1}^{n} \left | \left (Q_{l}(t) + \sum\limits_{j=0}^{k(n-d)-1}A_{l}(t+j) - \sum\limits_{j=0}^{k(n-d)-1}S_{l}(t+j) + \sum\limits_{j=0}^{k(n-d)-1}U_{l}(t+j) \right ) - Q_{l}(t) \right | \notag \\
    & = \sum\limits_{l=1}^{n} \left | \sum\limits_{j=0}^{k(n-d)-1}A_{l}(t+j) - \sum\limits_{j=0}^{k(n-d)-1}S_{l}(t+j) + \sum\limits_{j=0}^{k(n-d)-1}U_{l}(t+j) \right | \notag \\
    & \overset{(c)}{\leq} \sum\limits_{l=1}^{n} \left [ \left | \sum\limits_{j=0}^{k(n-d)-1}A_{l}(t+j) \right | + \left | \sum\limits_{j=0}^{k(n-d)-1}S_{l}(t+j) - \sum\limits_{j=0}^{k(n-d)-1}U_{l}(t+j) \right | \right ] \notag \\
    & \overset{(d)}{\leq} \sum\limits_{l=1}^{n} \left [ \sum\limits_{j=0}^{k(n-d)-1}A_{l}(t+j) + \sum\limits_{j=0}^{k(n-d)-1}S_{l}(t+j) \right ] \notag \\
    & \overset{(e)}{\leq} k(n-d)n(A_{\max}+S_{\max}), \label{eq:absolutebound}
\end{align}
where (a) follows from the fact that $\| \mathbf{x} \|_2 - \| \mathbf{y} \|_2 \leq \| \mathbf{x}-\mathbf{y} \|_2$ for any $\mathbf{x},\mathbf{y}\in\mathbb{R}^n $, (b) holds because $ \| \mathbf{x} \|_1 \geq \| \mathbf{x} \|_2 $, (c) follows from the triangle inequality, (d) follows from the fact that $ S_l(t) \geq U_l(t)\geq 0,\  \forall \ l,t $, and (e) is is due to the facts that $\|A(t)\|_1 \leq n A_{\max}$ and $S_l(t)\leq S_{\max}$ for all $l$ and $t$.\\

Secondly, we show that $W(\tilde{\mathbf{Q}}(t'))$ has a negative drift when the total queue length is sufficiently long, i.e., when  $ \| \tilde{\mathbf{Q}}(t') \|_1 \geq n k(n-d)S_{\max} $. Note that this assumption implies that there exists at least one queue $l$ such that $\tilde Q_l(t') \geq k(n-d)S_{\max} $. With this in mind, we divide the $n$ queues into a partition of five sets as follows. Given a fixed
\[ l'\in \underset{l\in [n]}{\arg\max} \left\{ \tilde{\mathbf{Q}}(t') \right\}, \]
and the set of the $d$ longest queues $\tilde I_d(t')$ selected by the algorithm, we define:
\begin{align}
    & I_1 := \{ l' \}, \notag \\
    & I_2:= \left \{ l \in \tilde I_d(t') \setminus \{ l^{\prime} \} : \tilde Q_l(t') \geq k(n-d)S_{\max} \right \}, \notag \\
    & I_3:= \left \{ l \in \tilde I_d(t') \setminus \{ l^{\prime} \} : \tilde Q_{l}(t') < k(n-d)S_{\max}  \right \}, \notag \\
    & I_4:= \left \{ l \in [n] \setminus \tilde I_d(t') : \tilde Q_{l}(t') \geq k(n-d)S_{\max} \right \}, \notag \\
    & I_5:= \left \{ l \in [n] \setminus \tilde I_d(t') : \tilde Q_{l}(t') < k(n-d)S_{\max} \right \}. \notag
\end{align}
Note that $I_1 \cup I_2 \cup I_3 = \tilde I_d(t')$, and that $I_4 \cup I_5 = [n] \setminus \tilde I_d(t')$. Using this partition, we bound the drift of $W(\tilde{\mathbf{Q}}(t'))$ as follows:
\begin{align}
    & E[\Delta W(\tilde{\mathbf{Q}}(t'))\ | \  \tilde{\mathbf{Q}}(t')=\mathbf{q}] \notag \\
    & = E[\| \tilde{\mathbf{Q}}(t'+1) \|_2^2 - \| \tilde{\mathbf{Q}}(t') \|_2^2 \  | \  \tilde{\mathbf{Q}}(t')=\mathbf{q} ] \notag \\
    & = E[\| \mathbf{Q}(t+k(n-d)) \|^2 - \| \mathbf{Q}(t) \|^2 \  | \ \mathbf{Q}(t)=\mathbf{q} ] \notag \\ 
    & = E \left [ \sum\limits_{l=1}^{n} \left (Q_{l}(t) + \sum\limits_{j=0}^{k(n-d)-1}A_{l}(t+j) - \sum\limits_{j=0}^{k(n-d)-1}S_{l}(t+j) + \sum\limits_{j=0}^{k(n-d)-1}U_{l}(t+j) \right )^2 - \sum\limits_{l=1}^{n} Q_{l}(t)^2 \  \middle | \  \mathbf{Q}(t)=\mathbf{q} \right ] \notag \\ 
    & = E \left [ \sum\limits_{l=1}^{n} \left [ \left (Q_{l}(t) + \sum\limits_{j=0}^{k(n-d)-1}A_{l}(t+j) - \sum\limits_{j=0}^{k(n-d)-1}S_{l}(t+j) + \sum\limits_{j=0}^{k(n-d)-1}U_{l}(t+j) \right )^2 - Q_{l}(t)^2  \right ] \  \middle | \ \mathbf{Q}(t)=\mathbf{q} \right] \notag \\ 
    & \overset{(a)}{=} E \left[ \left (Q_{l^{\prime}}(t) + \sum\limits_{j=0}^{k(n-d)-1}A_{l^{\prime}}(t+j) - \sum\limits_{j=0}^{k(n-d)-1}S_{l^{\prime}}(t+j) + \sum\limits_{j=0}^{k(n-d)-1}U_{l^{\prime}}(t+j) \right )^2 - Q_{l^{\prime}}(t)^2 \  \middle | \ \mathbf{Q}(t)=\mathbf{q} \right ] \notag \\
    & \quad + E \left[ \sum\limits_{ l\in I_2 } \left [ \left (Q_{l}(t) + \sum\limits_{j=0}^{k(n-d)-1}A_{l}(t+j) - \sum\limits_{j=0}^{k(n-d)-1}S_{l}(t+j) + \sum\limits_{j=0}^{k(n-d)-1}U_{l}(t+j) \right )^2 - Q_{l}(t)^2 \right ] \  \middle | \ \mathbf{Q}(t)=\mathbf{q} \right ] \notag \\
    & \quad + E \left[ \sum\limits_{ l\in I_3 } \left [ \left (Q_{l}(t) + \sum\limits_{j=0}^{k(n-d)-1}A_{l}(t+j) - \sum\limits_{j=0}^{k(n-d)-1}S_{l}(t+j) + \sum\limits_{j=0}^{k(n-d)-1}U_{l}(t+j) \right )^2 - Q_{l}(t)^2 \right ] \  \middle | \ \mathbf{Q}(t)=\mathbf{q} \right ] \notag \\
    & \quad + E \left[ \sum\limits_{ l\in I_4 } \left [ \left (Q_{l}(t) + \sum\limits_{j=0}^{k(n-d)-1}A_{l}(t+j) - \sum\limits_{j=0}^{k(n-d)-1}S_{l}(t+j) + \sum\limits_{j=0}^{k(n-d)-1}U_{l}(t+j)\right )^2 - Q_{l}(t)^2 \right ] \  \middle | \ \mathbf{Q}(t)=\mathbf{q} \right ] \notag \\
    & \quad + E \left[ \sum\limits_{ l \in I_5 } \left [ \left (Q_{l}(t) + \sum\limits_{j=0}^{k(n-d)-1}A_{l}(t+j) - \sum\limits_{j=0}^{k(n-d)-1}S_{l}(t+j) + \sum\limits_{j=0}^{k(n-d)-1}U_{l}(t+j) \right )^2 - Q_{l}(t)^2 \right ] \  \middle | \ \mathbf{Q}(t)=\mathbf{q} \right ] \notag \\
    & \overset{(b)}{\leq} E \left[ \left (Q_{l^{\prime}}(t) - \sum\limits_{j=0}^{k(n-d)-1}S_{l^{\prime}}(t+j) \right )^2 - Q_{l^{\prime}}(t)^2 \  \middle | \ \mathbf{Q}(t)=\mathbf{q} \right ] \notag \\
    & \quad + E \left[ \sum\limits_{ l\in I_2 } \left [ \left (Q_{l}(t) - \sum\limits_{j=0}^{k(n-d)-1}S_{l}(t+j) \right )^2 - Q_{l}(t)^2 \right ] \  \middle | \ \mathbf{Q}(t)=\mathbf{q} \right ] \notag \\
    & \quad + E \left[ \sum\limits_{ l\in I_4 } \left [ \left (Q_{l}(t) + \sum\limits_{j=0}^{k(n-d)-1}A_{l}(t+j) - \sum\limits_{j=0}^{k(n-d)-1}S_{l}(t+j) \right )^2 - Q_{l}(t)^2 \right ] \  \middle | \ \mathbf{Q}(t)=\mathbf{q} \right ] \notag \\
    & \quad + E \left[ \sum\limits_{ l \in I_5 } \left [ \left (Q_{l}(t) + \sum\limits_{j=0}^{k(n-d)-1}A_{l}(t+j) \right )^2 - Q_{l}(t)^2 \right ] \  \middle | \ \mathbf{Q}(t)=\mathbf{q} \right ], \label{eq:stable_expofwq}
\end{align}
where (a) follows from the definition of the partition, (b) follows from the following facts:
\begin{itemize}
    \item[(i)] The $d$ largest queues selected by the algorithm do not receive any arrivals during the $k(n-d)$ time slots, that is, we have $A_{l}(t+j) = 0$ for all $j \in \{ 0,...,k(n-d)-1\}$, for all $l\in I_1 \cup I_2 \cup I_3$.
    \item[(ii)] The queues which have at least $k(n-d)S_{\max}$ jobs in them at time $t$ will have no unused services for the $k(n-d)$ time slots, that is, we have $U_{l}(t+j)=0$ for all $j = \{ 0,...,k(n-d)-1 \}$ for all $l\in I_1 \cup I_2 \cup I_4$.
    \item[(iii)] The number of unused services is smaller than or equal to the number of potential services, that is, we have $ S_{l}(t+j) \geq U_{l}(t+j)$ for all $l\in[n]$ and $j\geq 0$.
\end{itemize}
We now compute the expectation in Equation \eqref{eq:stable_expofwq} for each term separately. For the longest queue $ l^{\prime} $, using the fact that potential services are independent from each other and from past queue lengths, we have
\begin{align}
    & E \left [ \left (Q_{l^{\prime}}(t)  - \sum\limits_{j=0}^{k(n-d)-1}S_{l^{\prime}}(t+j) \right )^2 - Q_{l^{\prime}}(t)^2 \  \middle | \  \mathbf{Q}(t)=\mathbf{q} \right ] \notag \\
    &\qquad\qquad\qquad = E \left [ -2 Q_{l^{\prime}}(t) \left ( \sum\limits_{j=0}^{k(n-d)-1}S_{l^{\prime}}(t+j) \right ) + \left ( \sum\limits_{j=0}^{k(n-d)-1}S_{l^{\prime}}(t+j) \right )^2 \  \middle | \  \mathbf{Q}(t)=\mathbf{q} \right ] \notag \\
    &\qquad\qquad\qquad = -2 q_{l^{\prime}}k(n-d)\mu_{l^{\prime}} + k^2(n-d)^2\mu_{l^{\prime}}^2+k(n-d)\sigma_{\mu_{l^{\prime}}}^2. \label{eq:stable_longest}
\end{align}
For the queues with indices in $ I_2 $, we have
\begin{align}
    & E \left [ \sum\limits_{ l \in I_2 } \left [ 
        \left (Q_{l}(t) - \sum\limits_{j=0}^{k(n-d)-1}S_{l}(t+j) \right )^2 - (Q_{l}(t))^2 
    \right ] \  \middle | \  \mathbf{Q}(t)=\mathbf{q} \right ] \notag \\
    & \qquad\qquad\qquad= E \left [ \sum\limits_{ l \in I_2 } \left [ 
        -2Q_{l}(t) \left (\sum\limits_{j=0}^{k(n-d)-1}S_{l}(t+j) \right ) + \left ( \sum\limits_{j=0}^{k(n-d)-1}S_{l}(t+j) \right )^2 
    \right ] \  \middle | \  \mathbf{Q}(t)=\mathbf{q} \right ] \notag \\
    & \qquad\qquad\qquad= \sum\limits_{ l \in I_2 } \left [
        -2 q_lk(n-d)\mu_l + k^2(n-d)^2\mu_l^2+k(n-d)\sigma_{\mu_l}^2
    \right ]. \label{eq:stable_I2}
\end{align}
For the queues with indices in $ I_4 $, we have
\begin{align}
    & E \left [ \sum\limits_{ l \in I_4 } \left [ 
        \left (Q_{l}(t) + \sum\limits_{j=0}^{k(n-d)-1}A_{l}(t+j) - \sum\limits_{j=0}^{k(n-d)-1}S_{l}(t+j) \right )^2 - (Q_{l}(t))^2 
    \right ] \  \middle | \  \mathbf{Q}(t)=\mathbf{q} \right ] \notag \\
    & \qquad\qquad = E \left [ \sum\limits_{ l \in I_4 } \left [ 
        2Q_{l}(t) \left (\sum\limits_{j=0}^{k(n-d)-1}A_{l}(t+j) - \sum\limits_{j=0}^{k(n-d)-1}S_{l}(t+j) \right )
    \right ] \  \middle | \  \mathbf{Q}(t)=\mathbf{q} \right ] \notag \\
    & \qquad\qquad\qquad\qquad + E \left [ \sum\limits_{ l \in I_4 } \left [ \left (\sum\limits_{j=0}^{k(n-d)-1}A_{l}(t+j) - \sum\limits_{j=0}^{k(n-d)-1}S_{l}(t+j) \right )^2 
    \right ] \  \middle | \  \mathbf{Q}(t)=\mathbf{q} \right ] \notag \\
    & \qquad\qquad = \sum\limits_{ l \in I_4 } \left [
        2q_{l} k \left ( n \lambda-(n-d)\mu_l \right )+ k^2 \left ( n \lambda-(n-d)\mu_l \right )^2 + kn\sigma_{\lambda}^2 +k(n-d)\sigma_{\mu_l}^2
    \right ]. \label{eq:stable_I4}
\end{align}
For queues with indices in $ I_5 $, using the fact that their queue lengths are bounded by $ k(n-d)S_{\max} $ by the definition of the partition, we have
\begin{align}
    & E \left [ \sum\limits_{ l \in I_5 } \left [ 
        \left (Q_{l}(t) + \sum\limits_{j=0}^{k(n-d)-1}A_{l}(t+j) \right )^2 - (Q_{l}(t))^2 
    \right ] \  \middle | \  \mathbf{Q}(t)=\mathbf{q} \right ] \notag \\
    & \qquad\qquad= E \left [ \sum\limits_{ l \in I_5 } \left [ 
        2Q_{l}(t) \left (\sum\limits_{j=0}^{k(n-d)-1}A_{l}(t+j) \right ) + \left (\sum\limits_{j=0}^{k(n-d)-1}A_{l}(t+j) \right )^2 
    \right ] \  \middle | \  \mathbf{Q}(t)=\mathbf{q} \right ] \notag \\
    & \qquad\qquad= \sum\limits_{ l\in I_5 } \left [
        2q_{l} k n\lambda + k^2 n^2\lambda^2 + kn\sigma_{\lambda}^2
    \right ] \notag \\
    & \qquad\qquad\leq \sum\limits_{ l\in I_5 } \left [
        2S_{\max}k^2(n-d) n\lambda + k^2 n^2\lambda^2 + kn\sigma_{\lambda}^2
    \right ] \label{eq:stable_I5}
\end{align}
Combining equations \eqref{eq:stable_expofwq}, \eqref{eq:stable_longest}, \eqref{eq:stable_I2}, \eqref{eq:stable_I4}, and \eqref{eq:stable_I5}, we get that
\begin{align}
    E[\Delta W(\tilde{\mathbf{Q}}(t'))\ | \  \tilde{\mathbf{Q}}(t')=\mathbf{q}] & = -2 q_{l^{\prime}}k(n-d)\mu_{l^{\prime}} + k^2(n-d)^2\mu_{l^{\prime}}^2+k(n-d)\sigma_{\mu_{l^{\prime}}}^2 \notag \\
    & \quad + \sum\limits_{ l \in I_2 } \left [ -2 q_l k (n-d)\mu_l + k^2(n-d)^2\mu_l^2+k(n-d)\sigma_{\mu_l}^2 \right ] \notag \\
    & \quad + \sum\limits_{ l \in I_4 } \left [ 2q_{l} k\left ( n\lambda-(n-d)\mu_l \right )+k^2\left ( n\lambda-(n-d)\mu_l \right )^2 + kn\sigma_{\lambda}^2 +k(n-d)\sigma_{\mu_l}^2 \right ] \notag \\
    & \quad + \sum\limits_{ l\in I_5 } \left [ 2q_l kn\lambda + k^2 n^2\lambda^2 + kn\sigma_{\lambda}^2 \right ] \notag \\
    & \leq -2 q_{l^{\prime}}k(n-d)\mu_{l^{\prime}} + k^2(n-d)^2\mu_{l^{\prime}}^2+k(n-d)\sigma_{\mu_{l^{\prime}}}^2 \notag \\
    & \quad + \sum\limits_{ l \in I_2 } \left [ -2 q_l k (n-d)\mu_l + k^2(n-d)^2\mu_l^2+k(n-d)\sigma_{\mu_l}^2 \right ] \notag \\
    & \quad + \sum\limits_{ l \in I_4 } \left [ 2q_{l} k\left ( n\lambda-(n-d)\mu_l \right )+k^2\left ( n\lambda-(n-d)\mu_l \right )^2 + kn\sigma_{\lambda}^2 +k(n-d)\sigma_{\mu_l}^2 \right ] \notag \\
    & \quad + \sum\limits_{ l\in I_5 } \left [ 2S_{\max}k(n-d) kn\lambda + k^2 n^2\lambda^2 + kn\sigma_{\lambda}^2 \right ] \notag \\
    & \overset{(a)}{=} - 2k(n-d)\mu_{l^{\prime}} q_{l^{\prime}} - \sum\limits_{ l\in I_2} 2k(n-d)\mu_l q_{l} + \sum\limits_{ l\in I_4} 2 k \left ( n\lambda-(n-d)\mu_l \right ) q_{l} \notag \\
    & \quad + k^2(n-d)^2\mu_{l^{\prime}}^2+k(n-d)\sigma_{\mu_{l^{\prime}}}^2 \notag \\
    & \quad + \sum\limits_{ l \in I_2 } \left [ k^2(n-d)^2\mu_l^2+k(n-d)\sigma_{\mu_l}^2 \right ] \notag \\
    & \quad + \sum\limits_{ l \in I_4 } \left [ k^2\left ( n\lambda-(n-d)\mu_l \right )^2 + kn\sigma_{\lambda}^2 +k(n-d)\sigma_{\mu}^2 \right ] \notag \\
    & \quad + \sum\limits_{ l\in I_5 } \left [ 2S_{\max}k^2(n-d)n\lambda + k^2n^2\lambda^2 + kn\sigma_{\lambda}^2 \right ] \notag \\
    & \overset{(b)}{\leq} - 2k(n-d)\mu_{l^{\prime}} q_{l^{\prime}} - \sum\limits_{ l\in I_2} 2k(n-d)\mu_l q_{l} + \sum\limits_{ l\in I_4} \left. 2 k \left ( n\lambda-(n-d)\mu_l \right ) q_{l} \right. \notag \\
    & \quad + k^2(n-d)^2\mu_{\max}^2+k(n-d)\sigma_{\max}^2 \notag \\
    & \quad + (d-1) \left [ k^2(n-d)^2\mu_{\max}^2+k(n-d)\sigma_{\max}^2 \right ] \notag \\
    & \quad + (n-d) \left [ k^2\left ( n\lambda + (n-d)\mu_{\max} \right )^2 + kn\sigma_{\lambda}^2 +k(n-d)\sigma_{\max}^2 \right ] \notag \\
    & \quad + (n-d) \left [ 2S_{\max}k^2(n-d)n\lambda + k^2 n^2\lambda^2 + kn\sigma_{\lambda}^2 \right ], \label{eq:stable_aveofwx}
\end{align}
where (a) follows from rearranging the terms, and (b) follows from the the definition of the partition, which yields $ |I_2|\leq d-1$, $|I_4| \leq n-d$, and $|I_5| \leq n-d $.

To further simplify the first term in Equation \eqref{eq:stable_aveofwx}, we define two subsets
\begin{align*}
    I_4^+ &:=\{ l\in I_4 : n\lambda-(n-d)\mu_l > 0 \}, \\
    I_4^- &:=\{ l\in I_4 : n\lambda-(n-d)\mu_l \leq 0 \}.
\end{align*}
We then have
\begin{align}
    & - 2k(n-d)\mu_{l^{\prime}} q_{l^{\prime}} - \sum\limits_{ l\in I_2} 2k(n-d)\mu_l q_{l} + \sum\limits_{ l\in I_4} 2k \left ( n\lambda - (n-d)\mu_l \right ) q_{l} \notag \\
    & = - 2k(n-d)\mu_{l^{\prime}} q_{l^{\prime}} - \sum\limits_{ l\in I_2} 2k(n-d)\mu_l q_{l} + \sum\limits_{ l\in I_4^+} 2k \left ( n\lambda - (n-d)\mu_l \right ) q_{l} + \sum\limits_{ l\in I_4^-} 2k \left ( n\lambda - (n-d)\mu_l \right ) q_{l} \notag \\
    & \overset{(a)}{\leq} - 2k(n-d)\mu_{l^{\prime}} q_{l^{\prime}} - \sum\limits_{ l\in I_2} 2k(n-d)\mu_l q_{l} + \sum\limits_{ l\in I_4^+} 2k \left ( n\lambda - (n-d)\mu_l \right ) q_{l}, \label{eq:stable_drift_I4}
\end{align}
where (a) follows from the definition of $ I_4^+$, and $I_4^-$. We have two different cases.

{\bf Case 1: $| I_4^+ | = 0$. } In this case, Equation \eqref{eq:stable_drift_I4} can be bounded as follows
\begin{align}
    & - 2k(n-d)\mu_{l^{\prime}} q_{l^{\prime}} - \sum\limits_{ l\in I_2} 2k(n-d)\mu_l q_{l} + \sum\limits_{ l\in I_4^+} 2k \left ( n\lambda - (n-d)\mu_l \right ) q_{l} \notag \\
    & = - 2k(n-d)\mu_{l^{\prime}} q_{l^{\prime}} - \sum\limits_{ l\in I_2} 2k(n-d)\mu_l q_{l} \notag \\
    &\leq - 2k(n-d)\mu_{\min} q_{l^{\prime}} \notag \\
    & \overset{(a)}{\leq} - \frac{2k(n-d)\mu_{\min}}{n}\| \mathbf{q} \|_1 \notag \\
    & \overset{(b)}{\leq} - \frac{2k(n-d)\mu_{\min}}{n}\| \mathbf{q} \|_2, \label{eq:stable_drift_case1}
\end{align} 
where  (a) follows from the fact that $ q_{l^{'}} \geq \| \mathbf{q} \|_1/n $, and (b) follows from the inequality $ \| \mathbf{x} \|_1 \geq \| \mathbf{x} \|_2 $ for all $ \mathbf{x} \in \mathbb{R}^n $.

{\bf Case 2: $| I_4^+ | > 0$.} In this case, we have $|I_3|=0$. Let
\[ l^{\prime\prime} \in  \argmin\limits_{ l \in I_2 } \{ q_l \} \qquad \text{and} \qquad l^{\prime\prime\prime} \in \underset{ l \in I_4^+}{\arg\max} \{ q_l \}, \]
and let
\begin{align}
    \epsilon^{\prime} & = \min_{I \subseteq [n] \ : \ |I| \geq d+1 } \  \left \{ \frac{1}{|I|-d} \sum_{ l\in I} \mu_l \right \} - \frac{n\lambda}{n-d}, \notag
\end{align} 
which is positive because we assumed that 
\[ n\lambda < \min_{I \subseteq [n] \ : \ |I| \geq d+1 } \  \left\{ \frac{n-d}{|I|-d} \sum_{ l\in I} \mu_l \right\}. \] 
Then, Equation \eqref{eq:stable_drift_I4} can be bounded as follows
\begin{align}
    & - 2k(n-d)\mu_{l^{\prime}} q_{l^{\prime}} - \sum\limits_{ l\in I_2} 2k(n-d)\mu_l q_{l} + \sum\limits_{ l\in I_4^+} 2k \left ( n\lambda - (n-d)\mu_l \right ) q_{l} \notag \\
    & \overset{(a)}{\leq} -2k(n-d)\mu_{l^{\prime}}q_{l^{\prime}} - 2k\sum\limits_{ l\in I_2} (n-d)\mu_l q_{l^{\prime\prime}} + 2k\sum\limits_{ l\in I_4^+} \left [ n\lambda - (n-d)\mu_l  \right ]q_{l^{\prime\prime\prime}} \notag \\
    & = -2k(n-d)\epsilon^{\prime}  q_{l^{\prime}} - 2k(n-d) \left [ (\mu_{l^{\prime}}-\epsilon^{\prime})q_{l^{\prime}} + \sum\limits_{ l\in I_2} \mu_l q_{l^{\prime\prime}} \right ] + 2k \sum\limits_{ l\in I_4^+} \left [ n\lambda-(n-d)\mu_l \right ]q_{l^{\prime\prime\prime}} \notag \\
    & \overset{(b)}{\leq} -2k(n-d)\epsilon^{\prime} q_{l^{\prime}} - 2k(n-d) \left ( \sum\limits_{ l\in I_1 \cup I_2} \mu_l - \epsilon^{\prime} \right ) q_{l^{\prime\prime}} + 2k \sum\limits_{ l\in I_4^+} n\lambda q_{l^{\prime\prime\prime}} -2k(n-d)\sum\limits_{ l\in I_4^+} \mu_lq_{l^{\prime\prime\prime}} \notag \\
    & \overset{(c)}{\leq} -2k(n-d)\epsilon^{\prime} q_{l^{\prime}} - 2k(n-d) \left ( \sum\limits_{ l\in I_1 \cup I_2} \mu_l - \epsilon^{\prime} \right ) q_{l^{\prime\prime\prime}} + 2k \sum\limits_{ l\in I_4^+} n\lambda q_{l^{\prime\prime\prime}} -2k(n-d)\sum\limits_{ l\in I_4^+} \mu_lq_{l^{\prime\prime\prime}} \notag \\
    & = -2k(n-d)\epsilon^{\prime} q_{l^{\prime}} - 2k(n-d) \left ( \sum\limits_{ l\in I_1 \cup I_2 \cup I_4^+} \mu_l - \epsilon^{\prime} \right ) q_{l^{\prime\prime\prime}} + 2k \sum\limits_{ l\in I_4^+} n\lambda q_{l^{\prime\prime\prime}} \notag \\
    & = -2k(n-d)\epsilon^{\prime} q_{l^{\prime}} - 2k \left [ (n-d) \left ( \sum\limits_{ l\in I_1 \cup I_2 \cup I_4^+} \mu_l - \epsilon^{\prime} \right ) - \sum\limits_{ l\in I_4^+} n\lambda \right ] q_{l^{\prime\prime\prime}}, \label{eq:stable_drift_case2}
\end{align} 
where (a) follows from the definitions of $ l^{\prime\prime} $ and $ l^{\prime\prime\prime} $, and (b) and (c) follow from the fact that $ q_{l^{\prime}} \geq q_{l^{\prime\prime}} \geq q_{l^{\prime\prime\prime}}$ by the definition of the partition. Since $I_4 \cup I_5 = [n]\backslash I_d(t)$ and $\emptyset \neq I_4^+ \subset I_4$, we have $| I_4^+ | = n-d-| I_5 |-|I_4^-| \in \{ 1,...,n-d\} $. Therefore, the last term in Equation \eqref{eq:stable_drift_case2} can be rewritten as follows:
\begin{align}
    (n-d) \left ( \sum\limits_{ l\in I_1 \cup I_2 \cup I_4^+} \mu_l - \epsilon^{\prime} \right ) - \sum\limits_{ l\in I_4^+} n\lambda & = (n-d) \sum\limits_{ l\in I_1 \cup I_2 \cup I_4^+} \mu_l - \sum\limits_{ l\in I_4^+}  n\lambda - (n-d)\epsilon^{\prime} \notag \\
    & = (n-d) \sum\limits_{ l\in I_1 \cup I_2 \cup I_4^+} \mu_l - (n-d-| I_5 |-|I_4^-|) n\lambda - (n-d)\epsilon^{\prime} \notag \\
    & = (n-d) \sum\limits_{ l\in I_1 \cup I_2 \cup I_4^+} \mu_l \notag \\ 
    & \quad - (n-d-| I_5 |-|I_4^-|) \left [ n\lambda + \frac{n-d}{(n-d-| I_5 |-|I_4^-|)} \epsilon^{\prime} \right ]. \label{eq:stable_case_I4+}
\end{align}
We now claim that this is positive. Let $ I_{m^*} $ be an optimal solution for the optimization problem that defines the capacity region, and let us reorder the service rates $ \{ \mu_1, \mu_2,...,\mu_n \} $ into a descending order $\mu_1^{\prime} \geq \mu_2^{\prime} \geq \cdots  \geq \mu_n^{\prime}$. We then have
\begin{align}
    n\lambda + \frac{n-d}{(n-d-| I_5 |-|I_4^-|)} \epsilon^{\prime} & < n\lambda + (n-d)\epsilon^{\prime} \notag \\
    & = \min_{I \subseteq [n] \ : \ |I| \geq d+1 } \  \left \{ \frac{n-d}{|I|-d} \sum_{ l\in I} \mu_l \right \} \notag \\
    & = \frac{n-d}{|I_{m^{*}}|-d}\sum\limits_{l=n-|I_{m^{*}}|+1}^{n} \mu_l^{\prime} \notag \\
    & \overset{(a)}{\leq} \frac{n-d}{n-| I_5 |-|I_4^-|-d} \sum\limits_{l=| I_5 |+|I_4^-|+1}^{n} \mu_l^{\prime} \notag \\ 
    & \overset{(b)}{\leq} \frac{n-d}{n-| I_5 |-|I_4^-|-d} \sum\limits_{l\in I_1 \cup I_2 \cup I_4^+} \mu_l, \notag
\end{align}
where (a) follows from the fact that $ n-| I_5 |-|I_4^-| \geq d+1$, and (b) follows from the fact that $ |I_1 \cup I_2 \cup I_4^+| = n-| I_5 |-|I_4^+|$ and $\mu_1^{\prime} \geq \mu_2^{\prime} \geq \cdots  \geq \mu_n^{\prime}$. It follows that \eqref{eq:stable_case_I4+} is positive, and thus Equation \eqref{eq:stable_drift_case2} can be further bounded by
\begin{align}
    -2k(n-d)\epsilon^{\prime} q_{l^{\prime}} - 2k \left [ (n-d) \left ( \sum\limits_{ l\in I_1 + I_2 + I_4^+} \mu_l - \epsilon^{\prime} \right ) - \sum\limits_{ l\in I_4^+} n\lambda \right ] q_{l^{\prime\prime\prime}}  & \leq - 2k(n-d)\epsilon^{\prime} q_{l^{\prime}} \notag \\
    & \leq - \frac{2\epsilon' (n-d)k}{n}\| \mathbf{q} \|_1 \notag \\
    & \leq - \frac{2\epsilon' (n-d)k}{n}\| \mathbf{q} \|_2. \notag 
\end{align}
Combining this with Equation \eqref{eq:stable_drift_case1}, we get that Equation \eqref{eq:stable_drift_I4} can be bounded as follows:
\begin{align}
    & - 2k(n-d)\mu_{l^{\prime}} q_{l^{\prime}} - \sum\limits_{ l\in I_2} 2k(n-d)\mu_l q_{l} + \sum\limits_{ l\in I_4} 2k \left ( n\lambda - (n-d)\mu_l \right ) q_{l} \leq - \frac{2 \min\{\mu_{\min},\,\epsilon'\} (n-d)k}{n}\| \mathbf{q} \|_2. \notag 
\end{align}
Finally, we have
\begin{align}
    E[\Delta W(\mathbf{Q}(t))\ | \  \mathbf{Q}(t)=\mathbf{q}]  & \leq - \frac{2 \min\{\mu_{\min},\,\epsilon'\} (n-d)k}{n}\| \mathbf{q} \|_2  + C_1 \label{eq:stable_wx}
\end{align}
where
\begin{align*}
     C_1 &= k^2(n-d)^2\mu_{\max}^2+k(n-d)\sigma_{\max}^2 + (d-1) \left [ k^2(n-d)^2\mu_{\max}^2+k(n-d)\sigma_{\max}^2 \right ] \notag \\
    & \quad \quad +  (n-d) \left [ k^2\left ( n\lambda + (n-d)\mu_{\max} \right )^2 + kn\sigma_{\lambda}^2 +k(n-d)\sigma_{\max}^2 \right ] \notag \\ 
    & \quad \quad + (n-d) \left [ 2S_{\max}k^2(n-d)n\lambda + k^2n^2\lambda^2 + kn\sigma_{\lambda}^2 \right ] \notag
\end{align*}
Since this drift is negative for all $\| \mathbf{q} \|_2$ large enough, the Foster-Lyapunov Theorem implies that the Markov chain $\{\tilde{\mathbf{X}}(t)\}_{t\geq 0} $ is positive recurrent. \\

We now prove the necessary condition. Suppose that
\begin{align}
    & n\lambda \geq \min_{I \subseteq [n] \ : \ |I| \geq d+1 } \  \left\{ \frac{n-d}{|I|-d} \sum_{ l\in I} \mu_l \right\}, \notag
\end{align}
and let $ I_{m^{*}} $ be an optimal solution. Let us reorder the service rates $\mu_1, \mu_2,...,\mu_n$ into a descending order $\mu_1^{\prime} \geq \mu_2^{\prime}\geq \cdots \geq \mu_n^{\prime}$. Let $\lambda_l$ be the average arrival rate to the $l$-th queue. Under the $k$-SLQ-$d$ policy, queues can receive at most one arrival every $n-d$ time slots. That is, we have $\lambda_l \leq n\lambda/(n-d)$, for all $l \in [n]$. Then, we have
\begin{align*}
    \sum\limits_{l=n-|I_{m^{*}}|+1}^n \mu_l^{\prime} - \sum\limits_{l=n-|I_{m^{*}}|+1}^{n}\lambda_l & = \frac{|I_{m^*}|-d}{n-d}\frac{n-d}{|I_{m^*}|-d} \sum\limits_{l=n-|I_{m^{*}}|+1}^n \mu_l^{\prime} - \sum\limits_{l=n-|I_{m^{*}}|+1}^{n}\lambda_l \\
    & \overset{(a)}{\leq} \frac{|I_{m^*}|-d}{n-d}n\lambda - \sum\limits_{l=n-|I_{m^{*}}|+1}^{n}\lambda_l \\
    & = \left ( 1 - \frac{n-|I_{m^*}|}{n-d} \right ) n \lambda - \sum\limits_{l=n-|I_{m^{*}}|+1}^{n}\lambda_l \\
    & \overset{(b)}{=} \sum\limits_{l=1}^{n-|I_{m^{*}}|} \lambda_l - \frac{n-|I_{m^*}|}{n-d}n\lambda \\
    & \leq 0,
\end{align*}
where (a) follows from the assumption above, and (b) follows from the fact that $ \sum_{l=1}^n \lambda_l = n\lambda $. Finally, note that the queue lengths can be lower bounded as
\begin{align}
    Q_{n-|I_{m^{*}}|+1}(t)+...+Q_n(t) & \geq \sum\limits_{i=1}^{t}\sum\limits_{l=n-|I_{m^{*}}|+1}^{n} (A_l[i] - S_l[i]), \notag
\end{align}
where the right hand side is a random walk with non-negative bias. Therefore, under both scenarios, the Markov chain $ \{\mathbf{X}(t) \}_{t\geq 0}$ cannot be positive recurrent.

\section{Proof of Corollary \ref{co_our:condition}} \label{prf:cor_stableregion}

First we prove the sufficient condition. It is enough to prove that, for any subset $I_m \subset [n]$ with $|I_m|\geq d+1$, we have
\begin{align}
    \frac{n-d}{|I_m|-d} \sum_{ l\in I_m} \mu_l \geq \sum\limits_{l=1}^{n} \mu_l. \notag 
\end{align}
Let $ I_m \subset [n]$, with $|I_m|=m \in \{d+1,...,n-1\}$, be the feasible solution for the objective function and reorder the service rates $\mu_1, \mu_2,...,\mu_n$ into a descending order $\mu_1^{\prime} \geq \mu_2^{\prime} \geq \cdots \geq \mu_n^{\prime}$. We have
\begin{align}
    \frac{n-d}{|I_m|-d} \sum_{ l\in I_m} \mu_l - \sum\limits_{l=1}^{n} \mu_l & \overset{(a)}{\geq} \frac{n-d}{|I_m|-d} \sum\limits_{l=n-|I_m|+1}^{n} \mu_l^{\prime} - \sum\limits_{l=1}^{n} \mu_l^{\prime} \notag \\
    & = \frac{n-d}{|I_m|-d} \left ( \sum\limits_{l=n-|I_m|+1}^{n} \mu_l^{\prime} + \sum\limits_{l=1}^{n-|I_m|} \mu_l^{\prime} - \sum\limits_{l=1}^{n-|I_m|} \mu_l^{\prime} \right ) - \sum\limits_{l=1}^{n} \mu_l^{\prime} \notag \\
    & = \frac{n-d}{|I_m|-d} \left ( \sum\limits_{l=1}^{n} \mu_l^{\prime} - \sum\limits_{l=1}^{n-|I_m|} \mu_l^{\prime} \right ) - \sum\limits_{l=1}^{n} \mu_l^{\prime} \notag \\
    & = \frac{n-|I_m|}{|I_m|-d}\sum\limits_{l=1}^{n} \mu_l^{\prime} - \frac{n-d}{|I_m|-d} \sum\limits_{l=1}^{n-|I_m|} \mu_l^{\prime} \notag \\ 
    & \overset{(b)}{\geq} \frac{n-|I_m|}{|I_m|-d}\sum\limits_{l=1}^{n} \mu_l^{\prime} - \frac{n-d}{|I_m|-d} (n-|I_m|) \mu_1^{\prime} \notag \\ 
    & = \frac{n-|I_m|}{|I_m|-d} \left [ \sum\limits_{l=1}^{n} \mu_l - (n-d) \mu_{\max} \right ] \notag \\
    & = \frac{n-m}{m-d} \left [ \sum\limits_{l=1}^{n} \mu_l - (n-d) \mu_{\max} \right ] \notag \\ 
    & \geq 0, \notag 
\end{align}
where (a) follows from the descending order of the service rates, and (b) follows from the fact that $ \mu_1^{\prime}=\mu_{\max} \geq \mu_l^{\prime}$, for all $l\geq 1$.\\

We now prove the necessary condition by contradiction. Suppose that the stability region is $n \lambda < \sum_{l=1}^{n} \mu_l $, and let us assume that $ \sum_{l=1}^{n} \mu_l < (n-d)\mu_{\max} $ holds. We define $ \epsilon := \sum_{l=1}^{n} \mu_l-n \lambda $, and reorder the service rates $\mu_1, \mu_2,...,\mu_n$ into a descending order $\mu_1^{\prime} \geq \mu_2^{\prime} \geq \cdots \geq \mu_n^{\prime}$. Note that the condition $ \sum_{l=1}^{n} \mu_l < (n-d)\mu_1' $ implies that
\begin{align}
    \frac{n-d}{n-d-1} \sum\limits_{l=2}^n \mu_l^{\prime} - \left ( \sum_{l=1}^n \mu_l^{\prime} - \epsilon \right ) & =  \frac{n-d}{n-d-1} \left ( \sum\limits_{l=2}^n \mu_l^{\prime} + \mu_1' -\mu_1' \right ) - \sum_{l=1}^n \mu_l^{\prime} + \epsilon \notag \\
    & = \frac{n-d}{n-d-1} \sum_{l=1}^n \mu_l^{\prime} - \sum_{l=1}^n \mu_l^{\prime} - \frac{n-d}{n-d-1} \mu_1' + \epsilon \notag \\
    & = \frac{1}{n-d-1} \sum_{l=1}^n \mu_l^{\prime} - \frac{n-d}{n-d-1} \mu_1' + \epsilon \notag \\
    & < 0, \label{eq:mu_for_small_epsilon}
\end{align}
for all $\epsilon$ small enough. Now let $ \lambda_l $ be the average arrival rate to the $l$-th queue. For all $\epsilon$ small enough, we have
\begin{align*}
    \sum\limits_{l=2}^n \mu_l^{\prime} - \sum\limits_{l=2}^{n}\lambda_l & = \frac{n-d-1}{n-d}\frac{n-d}{n-d-1} \sum\limits_{l=2}^n \mu_l^{\prime} - \sum\limits_{l=2}^{n}\lambda_l \\
    & \overset{(a)}{<} \frac{n-d-1}{n-d} \left (\sum_{l=1}^n \mu_l^{\prime} - \epsilon \right )  - \sum\limits_{l=2}^{n}\lambda_l \\
    & = \frac{n-d-1}{n-d} n\lambda  - \sum\limits_{l=2}^{n}\lambda_l \\
    & = \left ( 1 - \frac{1}{n-d} \right ) n \lambda - \sum\limits_{l=2}^{n}\lambda_l \\
    & = \lambda_1 - \frac{1}{n-d}n\lambda \\
    & \overset{(b)}{\leq} 0,
\end{align*}
where (a) follows from Equation \eqref{eq:mu_for_small_epsilon}, and (b) follows from the fact that under the $k$-SLQ-$d$ policy, queues receive at most one arrival every $n-d$ time slots, and thus $\lambda_l \leq n\lambda/(n-d)$, for all $l \in [n]$. It follows that
\begin{align}
    Q_2(t)+Q_3(t)+...+Q_n(t) & \geq \sum\limits_{i=1}^{t}\sum\limits_{l=2}^{n} (A_l[i] - S_l[i]), \notag 
\end{align}
where the right hand side is a random walk with positive bias. Therefore, the Markov chain $ \{\mathbf{X}(t) \}_{t\geq 0}$ is transient.

\section{Conclusion}
In this paper we introduce and analyze a new dispatching policy dubbed $k$-SLQ-$d$, which is throughput and delay optimal in classical and many-server heavy-traffic under certain conditions on the number of skipped servers $d$. Moreover, this policy can be implemented with an arbitrarily small communication overhead (by choosing the number of rounds $k$ to be sufficiently large) without sacrificing the throughput or delay optimality in heavy traffic. Moreover, our policy showcases how it is enough to skip the longest queue in order to obtain excellent delay performance, and that this can perform even better than sending jobs to the shortest queue when the communication overhead is extremely low.

\bibliographystyle{elsarticle-num.bst}
\bibliography{references}

\appendix
\newpage

\section{Proof of Theorem \ref{thm_our:SSC}} \label{prf:thm_SSC}
Let us define the Lyapunov Functions
\begin{align*}
    V_{\perp}(\mathbf{Q}(t)) &= \| \mathbf{Q}_{\perp}(t) \|_2, \\
    W(\mathbf{Q}(t)) &= \| \mathbf{Q}(t) \|_2^2, \\
    W_{\parallel}(\mathbf{Q}(t)) &= \| \mathbf{Q}_{\parallel}(t) \|_2^2.
\end{align*}
Given that the dispatcher collects the queue length vector every $k(n-d)$ time slots under our $k$-SLQ-$d$ policy, we define a new Markov chain 
\[ \tilde{\mathbf{X}}(t) := \Big( \tilde{\mathbf{Q}}(t), \tilde I_d(t), \tilde I_R(t), \tilde N_R(t)\Big) \]
such that
\[ \tilde{\mathbf{X}}(t)=\mathbf{X}(k(n-d)t) \]
for all $t\geq 0$. We also define the drift operator $\Delta$ by
\begin{align}
    \Delta V_{\perp}(\mathbf{Q}(t)) & := V(\mathbf{Q}_{\perp}(t+1))-V(\mathbf{Q}_{\perp}(t)) \notag \\
    \Delta W_{\parallel}(\mathbf{Q}(t)) & := W(\mathbf{Q}_{\parallel}(t+1))-W(\mathbf{Q}_{\parallel}(t)) \notag 
\end{align}

Fix $t\geq 0$ such that $t$ is a multiple of $k(n-d)$, and let $t'=t/k(n-d)$. We first show that $\Delta V_{\perp}(\tilde{\mathbf{Q}}(t'))$ is uniformly absolutely bounded, as follows.
\begin{align}
    | \Delta V_{\perp}(\tilde{\mathbf{Q}}(t')) | & = | \| \tilde{\mathbf{Q}}_{\perp}(t'+1) \|_2 - \| \tilde{\mathbf{Q}}_{\perp}(t') \|_2 | \notag \\
    & \overset{(a)}{\leq} \| \tilde{\mathbf{Q}}_{\perp}(t'+1) - \tilde{\mathbf{Q}}_{\perp}(t') \|_2 \notag \\
    & \overset{(b)}{=} \| \tilde{\mathbf{Q}}(t'+1) - \tilde{\mathbf{Q}}_{\parallel}(t'+1) - \tilde{\mathbf{Q}}(t') + \tilde{\mathbf{Q}}_{\parallel}(t')  \|_2 \notag \\
    & \overset{(c)}{\leq} \| \tilde{\mathbf{Q}}(t'+1) - \tilde{\mathbf{Q}}(t') \|_2 + \| \tilde{\mathbf{Q}}_{\parallel}(t'+1) - \tilde{\mathbf{Q}}_{\parallel}(t')  \|_2 \notag \\
    & \overset{(d)}{\leq} 2 \| \tilde{\mathbf{Q}}(t'+1) - \tilde{\mathbf{Q}}(t') \|_2 \notag \\
    & \overset{(e)}{\leq} 2 \| \tilde{\mathbf{Q}}(t'+1) - \tilde{\mathbf{Q}}(t') \|_1 \notag \\
    & = 2 \| \mathbf{Q}(t+k(n-d)) - \mathbf{Q}(t) \|_1 \notag \\
    & = 2 \sum\limits_{l=1}^{n} \left | \left (Q_{l}(t) + \sum\limits_{j=0}^{k(n-d)-1}A_{l}(t+j) - \sum\limits_{j=0}^{k(n-d)-1}S_{l}(t+j) + \sum\limits_{j=0}^{k(n-d)-1}U_{l}(t+j) \right ) - \left (Q_{l}(t) \right ) \right | \notag \\
    & = 2 \sum\limits_{l=1}^{n} \left | \sum\limits_{j=0}^{k(n-d)-1} A_{l}(t+j) - \sum\limits_{j=0}^{k(n-d)-1} S_{l}(t+j) + \sum\limits_{j=0}^{k(n-d)-1} U_{l}(t+j) \right | \notag \\
    & \leq 2 \sum\limits_{l=1}^{n} \left [ \left | \sum\limits_{j=0}^{k(n-d)-1} A_{l}(t+j) \right | + \left | \sum\limits_{j=0}^{k(n-d)-1} S_{l}(t+j) - \sum\limits_{j=0}^{k(n-d)-1} U_{l}(t+j) \right | \right ] \notag \\
    & \leq 2 \sum\limits_{l=1}^{n} \left [ \sum\limits_{j=0}^{k(n-d)-1} A_{l}(t+j) + \sum\limits_{j=0}^{k(n-d)-1} S_{l}(t+j)  \right ] \notag \\
    & \leq 2k(n-d)n(A_{\max}+S_{\max}), \notag
\end{align}
where (a) follows from the fact that $ | \| x \|_2 - \| y \|_2 | \leq \| x-y \|_2$ for $ x,y\in\mathbb{R}^L $, (b) follows from the fact that $ \mathbf{Q}_{\perp}(t) + \mathbf{Q}_{\parallel}(t) = \mathbf{Q}(t) $, (c) follows from the triangle inequality, (d) follows from the fact that $\mathbf{Q}_{\parallel}(t)$ is the projection of $\mathbf{Q}(t)$ onto $c$, which implies that $ \| \mathbf{Q}_{\parallel}(t'+1) - \mathbf{Q}_{\parallel}(t') \|_2 \leq \| \mathbf{Q}(t'+1) - \mathbf{Q}(t') \|_2 $, and (e) follows from the fact that $\| x \|_1 \geq \| x \|_2$. The remaining equalities and inequalities follow the same way as in the derivation of Equation \eqref{eq:absolutebound}.\\ 

We now show that $ \Delta V_{\perp}(\tilde{\mathbf{Q}}(t')) $ can be bounded by a function of $ \Delta W(\tilde{\mathbf{Q}}(t')) $ and $ \Delta W_{\parallel}(\tilde{\mathbf{Q}}(t')) $.
\begin{align}
    \Delta V_{\perp}(\tilde{\mathbf{Q}}(t')) & = \| \tilde{\mathbf{Q}}_{\perp}(t'+1) \|_2 - \| \tilde{\mathbf{Q}}_{\perp}(t') \|_2 \notag \\
    & = \sqrt{\| \tilde{\mathbf{Q}}_{\perp}(t'+1) \|_2^2} - \sqrt{\| \tilde{\mathbf{Q}}_{\perp}(t') \|_2^2} \notag \\
    & \overset{(a)}{\leq} \frac{1}{2\| \tilde{\mathbf{Q}}_{\perp}(t') \|_2}\left ( \| \tilde{\mathbf{Q}}_{\perp}(t'+1) \|_2^2 - \| \tilde{\mathbf{Q}}_{\perp}(t') \|_2^2 \right ) \notag \\
    & \overset{(b)}{=} \frac{1}{2\| \tilde{\mathbf{Q}}_{\perp}(t') \|_2}\left ( \| \tilde{\mathbf{Q}}(t'+1) \|_2^2 - \| \tilde{\mathbf{Q}}_{\parallel}(t'+1) \|_2^2 - \| \tilde{\mathbf{Q}}(t') \|_2^2 + \| \tilde{\mathbf{Q}}_{\parallel}(t') \|_2^2 \right ) \notag \\
    & = \frac{1}{2\| \tilde{\mathbf{Q}}_{\perp}(t') \|_2} \Big[\Delta W(\tilde{\mathbf{Q}}(t')) -\Delta W_{\parallel}(\tilde{\mathbf{Q}}(t')) \Big], \label{eq:difference_drifts}
\end{align}
where (a) follows from the fact that $ f(x)=\sqrt{x} $ is concave for $ x \geq 0 $, and thus $ f(y) - f(x)\leq (y-x)f^{\prime}(x) $, and (b) follows from the Pythagorean theorem.\\

We now proceed to bound the drift term $E[\Delta W(\tilde{\mathbf{Q}}(t'))\ | \ \tilde{\mathbf{Q}}(t') = \mathbf{q}] $. As it was shown in the proof of Theorem \ref{prf:thm_stable} (cf. Equation \eqref{eq:stable_aveofwx}), we have
\begin{align}
    & E\left[\Delta W\left(\tilde{\mathbf{Q}}(t')\right)\ | \ \tilde{\mathbf{Q}}(t')=\mathbf{q} \right] \notag \\ 
    & \leq -2 q_{l^{\prime}}k(n-d)\mu_{l^{\prime}} + k^2(n-d)^2\mu_{l^{\prime}}^2+k(n-d)\sigma_{\mu_{l^{\prime}}}^2 \notag \\
    & \quad + \sum\limits_{ l \in I_2 } \left [ -2 q_l k (n-d)\mu_l + k^2(n-d)^2\mu_l^2+k(n-d)\sigma_{\mu_l}^2 \right ] \notag \\
    & \quad + \sum\limits_{ l \in I_4 } \left [ 2q_{l} k\left ( \sum\limits_{i=1}^{n} \mu_i -\epsilon-(n-d)\mu_l \right )+k^2\left ( \sum\limits_{i=1}^{n}\mu_i - \epsilon-(n-d)\mu_l \right )^2 + kn\sigma_{\lambda}^2 +k(n-d)\sigma_{\mu_l}^2 \right ] \notag \\
    & \quad + \sum\limits_{ l\in I_5 } \left [ 2q_l k\left ( \sum\limits_{i=1}^{n}\mu_i -\epsilon \right ) + k^2\left ( \sum\limits_{i=1}^{n}\mu_i - \epsilon \right )^2 + kn \sigma_{\lambda}^2 \right ] \notag \\
    & = - 2k(n-d)\mu_{l^{\prime}} q_{l^{\prime}} - \sum\limits_{ l\in I_2} 2k(n-d)\mu_l q_{l} + \sum\limits_{ l\in I_4} 2 k \left ( \sum\limits_{i=1}^{n} \mu_i -\epsilon-(n-d)\mu_l \right ) q_{l} + \sum\limits_{ l\in I_5} 2 k \left ( \sum\limits_{i=1}^{n} \mu_i -\epsilon \right ) q_{l} \notag \\
    & \quad + k^2(n-d)^2\mu_{l^{\prime}}^2+k(n-d)\sigma_{\mu_{l^{\prime}}}^2 \notag \\
    & \quad + \sum\limits_{ l \in I_2 } \left [ -2 q_l k (n-d)\mu_l + k^2(n-d)^2\mu_l^2+k(n-d)\sigma_{\mu_l}^2 \right ] \notag \\
    & \quad + \sum\limits_{ l \in I_4 } \left [ k^2\left ( \sum\limits_{i=1}^{n}\mu_i - \epsilon-(n-d)\mu_l \right )^2 + kn\sigma_{\lambda}^2 +k(n-d)\sigma_{\mu_l}^2 \right ] \notag \\
    & \quad + \sum\limits_{ l\in I_5 } \left [ k^2\left ( \sum\limits_{i=1}^{n}\mu_i - \epsilon \right )^2 + kn \sigma_{\lambda}^2 \right ]. \label{eq:ssc_aveofssc}
\end{align}

The first term in Equation \eqref{eq:ssc_aveofssc} can be further bounded as follows:
\begin{align}
    & - 2k(n-d)\mu_{l^{\prime}} q_{l^{\prime}} - \sum\limits_{ l\in I_2} 2k(n-d)\mu_l q_{l} + \sum\limits_{ l\in I_4} 2k \left ( \sum\limits_{i=1}^{n} \mu_i -\epsilon-(n-d)\mu_l \right ) q_{l} + \sum\limits_{ l\in I_5} 2k \left ( \sum\limits_{i=1}^{n} \mu_i -\epsilon \right ) q_{l} \notag \\
    & = - 2k(n-d)\mu_{l^{\prime}} q_{l^{\prime}} -2k \sum\limits_{ l\in I_2 \cup I_3} (n-d)\mu_l q_{l} + 2k\sum\limits_{ l\in I_4 \cup I_5} \left ( \sum\limits_{i=1}^{n} \mu_i -\epsilon - (n-d)\mu_l \right ) q_{l} + 2k\sum\limits_{ l\in I_3 \cup I_5} \left ( n-d \right ) \mu_l q_{l} \notag \\ 
    & = - 2k(n-d) \epsilon \frac{ \| \mathbf{q} \|_1 }{n} - 2k(n-d) \left ( \mu_{l^{\prime}} - \frac{\epsilon}{n} \right) q_{l^{\prime}} -2k(n-d) \sum\limits_{ l\in I_2 \cup I_3} \left (\mu_l - \frac{\epsilon}{n} \right ) q_{l} \notag \\
    & \quad - 2k (n-d) \sum\limits_{ l\in I_4 \cup I_5} \left [ \mu_l - \left (\sum\limits_{i=1}^{n} \mu_i -\epsilon \right) \frac{1}{n-d} - \frac{\epsilon}{n} \right ] q_{l} + 2k (n-d) \sum\limits_{ l\in I_3 \cup I_5} \mu_l q_{l} \notag \\
    & = - \frac{2 \epsilon k(n-d)}{n} \| \mathbf{q} \|_1  - 2k(n-d) \sum\limits_{l=1}^n \beta_l q_l + 2k (n-d) \sum\limits_{ l\in I_3 \cup I_5} \mu_l q_{l}, \label{eq:ssc_defbeta}
\end{align}
where
\begin{align*}
    \beta_l = \left \{ \begin{array}{ll}
        \mu_l - \frac{\epsilon}{n}, & l \in \tilde{I_d}(t^{\prime}) \\
        \mu_l - \left (\sum\limits_{i=1}^{n} \mu_i -\epsilon \right) \frac{1}{n-d} - \frac{\epsilon}{n}, & l \in [n] \setminus \tilde{I_d}(t^{\prime})
    \end{array} \right.
\end{align*}
In particular, note that $\sum_{l=1}^n \beta_l = 0$. Equation \eqref{eq:ssc_defbeta} can be rearranged as follows:
\begin{align}
    & - \frac{2 \epsilon k(n-d)}{n} \| \mathbf{q} \|_1  - 2k(n-d) \sum\limits_{l=1}^n \beta_l q_l + 2k (n-d) \sum\limits_{ l\in I_3 \cup I_5} \mu_l q_{l} \notag \\
    & = - \frac{2 \epsilon k(n-d)}{n} \| \mathbf{q} \|_1 - 2k(n-d)\beta_1(q_1-q_2) - 2k(n-d) \left ( \sum\limits_{r=1}^2 \beta_r \right ) (q_2 - q_3) \notag \\ 
    & \quad - ... - 2k(n-d) \left ( \sum\limits_{r=1}^{n-1} \beta_r \right ) (q_{n-1} - q_n) - 2k(n-d) \left ( \sum\limits_{r=1}^{n} \beta_r \right ) q_n + 2k (n-d) \sum\limits_{ l\in I_3 \cup I_5} \mu_l q_{l} \notag \\
    & = - \frac{2 \epsilon k(n-d)}{n} \| \mathbf{q} \|_1 - 2k(n-d) \sum\limits_{l=1}^{n-1} \left [ \left( \sum\limits_{r=1}^{l} \beta_r \right )(q_l - q_{l+1}) \right ] + 2k (n-d) \sum\limits_{ l\in I_3 \cup I_5} \mu_l q_{l} \label{eq:ssc_sumbeta}
\end{align}
We now bound $ \sum_{r=1}^l \beta_r $ for two different cases.\\

{\bf Case 1:} If $ 1\leq l \leq d$, we have:
\begin{align*}
    \sum_{r=1}^l \beta_r & = \sum\limits_{r=1}^l \left( \mu_r - \frac{\epsilon}{n} \right) \\
    & \geq \sum\limits_{r=1}^l \mu_r - \epsilon \\
    & \geq \mu_{\min} - \epsilon \\
    & > 0
\end{align*}
Where in the last inequality we used that $\epsilon < \mu_{\min}$.\\

{\bf Case 2:} If $d+1\leq l \leq n-1 $, we have:
\begin{align*}
    \sum_{r=1}^l \beta_r & = \sum\limits_{r=1}^d \left( \mu_r - \frac{\epsilon}{n} \right) + \sum\limits_{r=d+1}^l \left( \mu_r - \left( \sum\limits_{i=1}^n \mu_i - \epsilon \right) \frac{1}{n-d} - \frac{\epsilon}{n} \right) \\
    & = \sum\limits_{r=1}^l \mu_r - \left ( \sum\limits_{i=1}^n \mu_i \right ) \frac{l-d}{n-d} + \epsilon \left ( \frac{l-d}{n-d} - \frac{l}{n} \right ) \\
    & = \sum\limits_{i=1}^n \mu_i \left ( \frac{\sum\limits_{r=1}^l \mu_r }{\sum\limits_{i=1}^n \mu_i} - \frac{l-d}{n-d} \right ) + \epsilon \left ( \frac{l-d}{n-d} - \frac{l}{n} \right ) \\
    & = \sum\limits_{i=1}^n \mu_i \left [ 1 - \frac{\sum\limits_{r=l+1}^n \mu_r }{\sum\limits_{i=1}^n \mu_i} - \left ( 1 - \frac{n-l}{n-d} \right ) \right ] + \epsilon \left ( \frac{l-d}{n-d} - \frac{l}{n} \right ) \\
    & = \sum\limits_{i=1}^n \mu_i \left ( \frac{n-l}{n-d} - \frac{\sum\limits_{r=l+1}^n \mu_r }{\sum\limits_{i=1}^n \mu_i} \right ) + \epsilon \left ( \frac{l-d}{n-d} - \frac{l}{n} \right ) \\
    & \geq \sum\limits_{i=1}^n \mu_i \left ( \frac{n-l}{n-d} - \frac{ (n-l) \mu_{max} }{\sum\limits_{i=1}^n \mu_i} \right ) + \epsilon \left ( \frac{l-d}{n-d} - \frac{l}{n} \right ) \\
    & \geq \sum\limits_{i=1}^n \mu_i \left ( \frac{n-l}{n-d} - \frac{ (n-l) \mu_{max} }{\sum\limits_{i=1}^n \mu_i} \right ) - \epsilon \\
    & \overset{(1)}{\geq} \sum\limits_{i=1}^n \mu_i \left ( \frac{1}{n-d} - \frac{ \mu_{max} }{\sum\limits_{i=1}^n \mu_i} \right ) - \epsilon \\
    & = \frac{1}{n-d} \sum\limits_{i=1}^n \mu_i - \mu_{\max} - \epsilon \\
    & \overset{(2)}{>} 0,
\end{align*}
where (1) follows from Equation \eqref{eq:d_large_enough}, and (2) follows from the fact that $\epsilon < \sum_{i=1}^n \mu_i / (n-d) - \mu_{\max}$.\\

Let $\epsilon$ be such that
\[ \epsilon < \frac{1}{2}\min \left \{ \mu_{\min}, \left ( \frac{1}{n-d} \sum\limits_{i=1}^n \mu_i \right ) - \mu_{\max} \right \}. \]
Then, Equation \eqref{eq:ssc_sumbeta} can be further bounded as follows:
\begin{align} 
    & - \frac{2 \epsilon k(n-d)}{n} \| \mathbf{q} \|_1 - 2k(n-d) \sum\limits_{l=1}^{n-1} \left [ \left( \sum\limits_{r=1}^{l} \beta_r \right )(q_l - q_{l+1}) \right ] + 2k (n-d) \sum\limits_{ l\in I_3 \cup I_5} \mu_l q_{l} \notag \\
    & \overset{(a)}{\leq} - \frac{2 \epsilon k(n-d)}{n} \| \mathbf{q} \|_1 -2k(n-d) \left ( \min \left \{ \mu_{\min}, \left ( \frac{1}{n-d} \sum\limits_{i=1}^n \mu_i \right ) - \mu_{\max} \right \} - \epsilon \right )(q_1 - q_n) + 2k (n-d) \sum\limits_{ l\in I_3 \cup I_5} \mu_l q_{l} \notag \\
    & \overset{(b)}{\leq} - \frac{ 2 \epsilon k(n-d)}{\sqrt{n}} \| \mathbf{q_{\parallel}} \|_2 -2k(n-d) \left ( \min \left \{ \mu_{\min}, \left ( \frac{1}{n-d} \sum\limits_{i=1}^n \mu_i \right ) - \mu_{\max} \right \} - \epsilon \right )(q_1 - q_n) + 2k (n-d) \sum\limits_{ l\in I_3 \cup I_5} \mu_l q_{l} \notag \\
    & \overset{(c)}{\leq} - \frac{ 2 \epsilon k(n-d)}{\sqrt{n}} \| \mathbf{q_{\parallel}} \|_2 - \frac{2k(n-d) \left ( \min \left \{ \mu_{\min}, \left ( \frac{1}{n-d} \sum\limits_{i=1}^n \mu_i \right ) - \mu_{\max} \right \} - \epsilon \right )}{\sqrt{n}} \| \mathbf{q}_{\perp} \|_2 + 2k (n-d) \sum\limits_{ l\in I_3 \cup I_5} \mu_l q_{l} \notag \\
    & \overset{(d)}{\leq} - \frac{ 2 \epsilon k(n-d)}{\sqrt{n}}v \| \mathbf{q_{\parallel}} \|_2 - \frac{2k(n-d) \left ( \min \left \{ \mu_{\min}, \left ( \frac{1}{n-d} \sum\limits_{i=1}^n \mu_i \right ) - \mu_{\max} \right \} - \epsilon \right )}{\sqrt{n}} \| \mathbf{q}_{\perp} \|_2 \notag \\
    & \quad + 2k^2(n-1)(n-d)^2 \mu_{\max}S_{\max}, \label{eq:SSC_waytodrift}
\end{align}
where (a) follows from the definition of $\epsilon$, (b) follows from the fact that $ \| \mathbf{q} \|_1/n = \sqrt{n(\mathbf{q}_{\Sigma}(t)/n)^2}/\sqrt{n} = \| \mathbf{q}_{\parallel} \|_2/\sqrt{n} $, (c) follows from the definition of $\| q_{\perp} \|_2$ and the fact that $q_1 \geq q_l$ for all $l \in [n]$ and $ (\sum_{s=1}^n q_s) / n \geq q_1$, which implies
\begin{align*}
    \| \mathbf{q}_{\perp} \|_2^2 = \sum\limits_{l=1}^n \left ( q_l -\frac{\sum\limits_{s=1}^n q_s}{n} \right )^2 \leq n (q_1 - q_n)^2,
\end{align*} 
and (d) follows from the fact that $ |I_3|+| I_5 | \leq n-1 $ and $ q_l \leq k(n-d)S_{\max}$ for all $l \in I_3 \cup I_5 $.\\

Combining equations \eqref{eq:ssc_aveofssc} and \eqref{eq:SSC_waytodrift}, we get
\begin{align}
    & E[\Delta W(\tilde{\mathbf{Q}}(t'))\ | \ \tilde{\mathbf{Q}}(t')=q] \notag \\
    & \leq - 2k(n-d)\mu_{l^{\prime}} q_{l^{\prime}} - \sum\limits_{ l\in I_2} 2k(n-d)\mu_l q_{l} + \sum\limits_{ l\in I_4} 2 k \left ( \sum\limits_{l=1}^{n} \mu_l -\epsilon-(n-d)\mu_l \right ) q_{l} + \sum\limits_{ l\in I_5} 2 k \left ( \sum\limits_{l=1}^{n} \mu_l -\epsilon \right ) q_{l} \notag \\
    & \quad + k^2(n-d)^2\mu_{\max}^2+k(n-d)\sigma_{\mu_{\max}}^2 \notag \\
    & \quad + (d-1) \left [ k^2(n-d)^2\mu_{\max}^2+k(n-d)\sigma_{\mu_{\max}}^2 \right ] \notag \\
    & \quad + (n-d) \left [ k^2\left ( \sum\limits_{l=1}^{n} \mu_l - \epsilon + (n-d)\mu_{\max} \right )^2 + kn \sigma_{\lambda}^2 +k(n-d)\sigma_{\mu_{\max}}^2 \right ] \notag \\
    & \quad + (n-d) \left [ k^2 \left ( \sum\limits_{l=1}^{n} \mu_l-\epsilon \right )^2 + kn \sigma_{\lambda}^2 \right ] \notag \\
    & \leq - \frac{ 2\epsilon k(n-d)}{\sqrt{n}} \| \mathbf{q}_{\parallel} \|_2 - \frac{2k(n-d) \left ( \min \left \{ \mu_{\min}, \left ( \frac{1}{n-d} \sum\limits_{i=1}^n \mu_i \right) - \mu_{\max} \right \} - \epsilon \right )}{\sqrt{n}} \| \mathbf{q}_{\perp} \|_2 + C_1^{\prime}, \label{eq:ssc_driftofssc} 
\end{align}
where
\begin{align}
    & C_1^{\prime} = 2k^2(n-1)(n-d)^2\mu_{\max} S_{\max} + k^2(n-d)^2\mu_{\max}^2+k(n-d)\sigma_{\mu_{\max}}^2 \notag \\
    & \qquad+ (d-1) \left [ k^2(n-d)^2\mu_{\max}^2 + k(n-d)\sigma_{\mu_{\max}}^2 \right ] \notag \\
    & \qquad + (n-d) \left [ k^2\left ( \sum\limits_{l=1}^{n} \mu_l - \epsilon + (n-d)\mu_{\max} \right )^2 + kn \sigma_{\lambda}^2 +k(n-d)\sigma_{\mu_{\max}}^2 \right ] + (n-d) \left [ k^2 \left ( \sum\limits_{l=1}^{n} \mu_l-\epsilon \right )^2 + kn \sigma_{\lambda}^2 \right ]. \notag 
\end{align}

We now lower bound $ E[\Delta W_{\parallel}(\tilde{\mathbf{Q}}(t'))\ | \  \tilde{\mathbf{Q}}(t')=\mathbf{q}] $ as follows:
\begin{align}
    & E[\Delta W_{\parallel}(\tilde{\mathbf{Q}}(t'))\ | \ \tilde{\mathbf{Q}}(t') = \mathbf{q}] = E[\| \tilde{\mathbf{Q}}_{\parallel}(t'+1) \|_2^2 - \| \tilde{\mathbf{Q}}_{\parallel}(t') \|_2^2 \  | \  \tilde{\mathbf{Q}}(t')=\mathbf{q} ] \notag \\
    & = E[\| \mathbf{Q}_{\parallel}(t+k(n-d)) \|_2^2 - \| \mathbf{Q}_{\parallel}(t) \|_2^2 \  | \  \mathbf{Q}(t)=\mathbf{q} ] \notag \\
    & = E \left [ \sum\limits_{l=1}^{n} \left ( \frac{Q_{\Sigma}(t+k(n-d))}{n} \right )^2 - \sum\limits_{l=1}^{n} \left ( \frac{Q_{\Sigma}(t)}{n} \right )^2 \ \middle | \  \mathbf{Q}(t)=\mathbf{q} \right ] \notag \\ 
    & = E \left [ \frac{1}{n}(Q_{\Sigma}(t+k(n-d))^2 - Q_{\Sigma}(t)^2) \  \middle | \  \mathbf{Q}(t)=\mathbf{q} \right ] \notag \\
    & = \frac{1}{n} E \left [ \left ( Q_{\Sigma}(t) + \sum\limits_{j=0}^{k(n-d)-1} \sum\limits_{l=1}^n A_l(t+j) - \sum\limits_{j=0}^{k(n-d)-1} \sum\limits_{l=1}^n S_l(t+j) + \sum\limits_{j=0}^{k(n-d)-1} \sum\limits_{l=1}^n U_l(t+j) \right )^2 - Q_{\Sigma}(t)^2 \  \middle | \  \mathbf{Q}(t)=\mathbf{q} \right ] \notag \\
    & = \frac{1}{n} \left \{ E \left [ \left ( Q_{\Sigma}(t) + \sum\limits_{j=0}^{n-d-1} \sum\limits_{l=1}^n A_l(t+j) - \sum\limits_{j=0}^{k(n-d)-1} \sum\limits_{l=1}^n S_l(t+j) \right )^2 - Q_{\Sigma}(t)^2 \  \middle | \  \mathbf{Q}(t)=\mathbf{q} \right ] \right. \notag \\
    & \quad + E \left [ 2\left ( Q_{\Sigma}(t) + \sum\limits_{j=0}^{k(n-d)-1} \sum\limits_{l=1}^n A_l(t+j) - \sum\limits_{j=0}^{k(n-d)-1} \sum\limits_{l=1}^n S_l(t+j) \right )\left ( \sum\limits_{j=0}^{k(n-d)-1} \sum\limits_{l=1}^n U_l(t+j) \right) \  \middle | \  \mathbf{Q}(t)=\mathbf{q} \right ] \notag \\
    & \quad \left. + E \left [ \left ( \sum\limits_{j=0}^{k(n-d)-1} \sum\limits_{l=1}^n U_l(t+j) \right)^2 \  \middle | \  \mathbf{Q}(t)=\mathbf{q} \right ] \right \} \notag \\
    & = \frac{1}{n} \left \{ E \left [ 2 Q_{\Sigma}(t) \left (\sum\limits_{j=0}^{k(n-d)-1} \sum\limits_{l=1}^n A_l(t+j) - \sum\limits_{j=0}^{k(n-d)-1} \sum\limits_{l=1}^n S_l(t+j) \right ) \  \middle | \  \mathbf{Q}(t)=\mathbf{q} \right ] \right. \notag \\
    & \quad + E \left [ \left (\sum\limits_{j=0}^{k(n-d)-1} \sum\limits_{l=1}^n A_l(t+j) - \sum\limits_{j=0}^{k(n-d)-1} \sum\limits_{l=1}^n S_l(t+j) \right ) ^2 \  \middle | \  \mathbf{Q}(t)=\mathbf{q} \right ] \notag \\
    & \quad + E \left [ 2 \left ( Q_{\Sigma}(t) + \sum\limits_{j=0}^{k(n-d)-1} \sum\limits_{l=1}^n A_l(t+j) \right )\left ( \sum\limits_{j=0}^{k(n-d)-1} \sum\limits_{l=1}^n U_l(t+j) \right) \  \middle | \  \mathbf{Q}(t)=\mathbf{q} \right ] \notag \\
    & \quad - E \left [ 2\left ( \sum\limits_{j=0}^{k(n-d)-1} \sum\limits_{l=1}^n S_l(t+j) \right) \left ( \sum\limits_{j=0}^{k(n-d)-1} \sum\limits_{l=1}^n U_l(t+j) \right) \  \middle | \  \mathbf{Q}(t)=\mathbf{q} \right ] \notag \\
    & \quad \left. + E \left [ + \left ( \sum\limits_{j=0}^{k(n-d)-1} \sum\limits_{l=1}^n U_l(t+j) \right)^2 \  \middle | \  \mathbf{Q}(t)=\mathbf{q} \right ] \right \} \notag \\
    & \overset{(a)}{\geq} \frac{1}{n} \left \{ E \left[ 2 Q_{\Sigma}(t) \left (\sum\limits_{j=0}^{k(n-d)-1} \sum\limits_{l=1}^n A_l(t+j) - \sum\limits_{j=0}^{k(n-d)-1} \sum\limits_{l=1}^n S_l(t+j) \right ) \  \middle | \  \mathbf{Q}(t)=\mathbf{q} \right ] \right. \notag \\
    & \quad \left. - E \left [ 2\left ( \sum\limits_{j=0}^{k(n-d)-1} \sum\limits_{l=1}^n S_l(t+j) \right) \left ( \sum\limits_{j=0}^{k(n-d)-1} \sum\limits_{l=1}^n U_l(t+j) \right) \  \middle | \  \mathbf{Q}(t)=\mathbf{q} \right ] \right \} \notag \\
    & \overset{(b)}{\geq} 2\frac{1}{n} \left ( k(n-d)\left (\sum\limits_{l=1}^{n}\mu_l-\epsilon \right )-k(n-d)\sum\limits_{l=1}^{n}\mu_l \right )q_{\Sigma}(t) - 2\frac{1}{n}k(n-d)nS_{\max}k(n-d)nS_{\max} \notag \\
    & = - \frac{2 \epsilon k(n-d)}{n}\|\mathbf{q} \|_1 - 2 k^2 (n-d)^2 n S_{\max}^2 \notag \\
    & = - \frac{2 \epsilon k(n-d)}{\sqrt{n}}\| \mathbf{q}_{\parallel} \|_2 - C_2, \label{eq:ssc_dirftofparallel}
\end{align}
with $C_2 = 2k^2(n-d)^2 n S_{\max}^2$, where (a) follows from removing the positive term, and (b) follows from using that $ U_l(t+j) \leq S_{\max}$ for all $l \in [n]$ and $j\geq 0$. \\

Finally, combining equations \eqref{eq:difference_drifts}, \eqref{eq:ssc_driftofssc}, and \eqref{eq:ssc_dirftofparallel}, we get
\begin{align}
         & E[ \Delta V_{\perp}(\tilde{\mathbf{Q}}(t'))\  |\ \tilde{\mathbf{Q}}(t')=\mathbf{q} ] \notag \\ 
         & \leq \frac{1}{2 \| \mathbf{q}_{\perp} \|_2 } E \left [\Delta W(\tilde{\mathbf{Q}}(t')) - \Delta W_{\parallel}(\tilde{\mathbf{Q}}(t')) \ \middle |\ \tilde{\mathbf{Q}}(t')= \mathbf{q} \right ] \notag \\ 
         & = \frac{1}{2 \| \mathbf{q}_{\perp} \|_2 } \Bigg [ -2\epsilon k \frac{n-d}{\sqrt{n}}\| \mathbf{q}_{\parallel} \|_2 - \frac{2k(n-d) \left ( \min \left \{ \mu_{\min}, \left ( \frac{1}{n-d} \sum\limits_{i=1}^n \mu_i \right ) - \mu_{\max} \right \} - \epsilon \right )}{\sqrt{n}} \left \| \mathbf{q}_{\perp} \right \|_2 + C_1^{\prime} \notag \\
         & \quad + 2\epsilon k \frac{n-d}{\sqrt{n}}\| \mathbf{q}_{\parallel} \|_2 + C_2 \Bigg ]  \notag \\
         & = - \frac{ k(n-d) \left ( \min \left \{ \mu_{\min}, \left (\frac{1}{n-d} \sum\limits_{i=1}^n \mu_i \right ) - \mu_{\max} \right \} - \epsilon \right )}{\sqrt{n}} + \frac{C}{2 \| \mathbf{q}_{\perp} \|_2 }. \notag 
\end{align}
In particular, for
\[ \epsilon \leq \frac{1}{2}\min \left \{ \mu_{\min}, \left ( \frac{1}{n-d} \sum\limits_{i=1}^n \mu_i \right ) - \mu_{\max} \right \}, \]
this implies that 
\begin{align}
    E \left [ \Delta V_{\perp}(\tilde{\mathbf{Q}}_{\perp}) + \epsilon_0 \  \middle | \  \| \tilde{\mathbf{Q}}_{\perp} \|_2 > a \right ] < 0, \notag
\end{align}
for
\begin{align*}
     a &= \frac{C \sqrt{n} }{k (n-d) \Delta } \qquad \text{and} \qquad \epsilon_0 = \frac{k (n-d) \Delta}{2 \sqrt{n} },
\end{align*}
where
\[ \Delta = \frac{\min \left \{ \mu_{\min}, \left ( \frac{1}{n-d} \sum\limits_{i=1}^n \mu_i \right ) - \mu_{\max} \right \}}{2}. \]
Then, Theorem 2.3 in \cite{hajek1982hitting-lyapunovupperbound} implies that
\begin{align}
    & E \left [ e^{\eta\| \tilde{\mathbf{Q}}_{\perp}\|_2} \right ] \leq \frac{1}{1-\rho}e^{\eta a + 1}, \label{eq:exp_bound}
\end{align}
where
\begin{align}
    & \eta = \min\left\{ \frac{1}{Z}, \frac{k (n-d) \Delta}{4 \sqrt{n} Z^2(e-2)}, \frac{k (n-d) \Delta}{C \sqrt{n}} \right\} \notag \\
    & \rho = 1 - \epsilon_0\eta + Z^2(e-2)\eta^2, \notag
\end{align}
with
\[ Z = 2k(n-d)n(A_{\max}+S_{\max}). \]
Taking the Taylor expansion for the left-hand side in Equation \eqref{eq:exp_bound} yields
\[ E \left [ 1 + \eta \| \tilde{\mathbf{Q}}_{\perp} \|_2 + \frac{\eta^2}{2!}\| \tilde{\mathbf{Q}}_{\perp} \|_2^2 + \sum\limits_{k=3}^{\infty}\frac{\eta^k}{k!} \| \tilde{\mathbf{Q}}_{\perp} \|_2^k \right ] \leq \frac{1}{1-\rho}e^{\eta a + 1}, \]
and thus
\begin{align}
    E \left [ \| \tilde{\mathbf{Q}}_{\perp} \|_2^2 \right ] &\leq \frac{2}{\eta^2}\frac{1}{1-\rho}e^{\eta a+1} \notag \\
    & = \frac{4 \sqrt{n} e^{\eta C \sqrt{n} /(k (n-d) \Delta)+1}}{k (n-d) \Delta \eta^3 - 2 \sqrt{n} Z^2 (e-2)\eta^4} \notag \\
    &= \frac{4 \sqrt{n} e^2}{k (n-d) \Delta \eta^3 - 2 \sqrt{n} Z^2(e-2)\eta^4} \notag \\ 
    & =: N_2(n,k,d). \notag 
\end{align}

Finally, even though the State Space Collapse result is only for the sampled queue length vector $\tilde{\mathbf{Q}}$, in between those sampling points the queues can only increase or decrease by at most $k(nA_{\max}+(n-d)S_{\max})$. Therefore, $\|\tilde{\mathbf{Q}}_{\perp}\|_2^2$ can only increase by at most $k^2(nA_{\max}+(n-d)S_{\max})^2n$, which is constant in $\epsilon$ and of order $\Theta(k^2n^3)$, which is smaller than or equal to $N_2$. It follows that we have the same State Space Collapse result for $\mathbf{Q}$.

\section{Proof of Theorem \ref{thm_our:SSCUpper}} \label{prf:thm_upperbound}
Suppose that $\tilde{\mathbf{Q}}(t')=\mathbf{Q}(t)$ is in steady-state. Then, we have
\begin{align}
    & E \left [ \Delta W_{\parallel} ( \tilde{\mathbf{Q}}(t') ) \right ] = E \left [ \| \tilde{\mathbf{Q}}_{\parallel}(t'+1) \|^2 - \| \tilde{\mathbf{Q}}_{\parallel}(t') \|^2 \right ] \notag \\
    & = E\left [ \| \mathbf{Q}_{\parallel}(t+k(n-d)) \|^2 - \| \mathbf{Q}_{\parallel}(t) \|^2 \right ] \notag \\
    & = E \left [ \sum\limits_{l=1}^{n} \left ( \frac{Q_{\Sigma}(t+k(n-d))}{n} \right )^2 - \sum\limits_{l=1}^{n} \left ( \frac{Q_{\Sigma}(t)}{n} \right )^2 \right ] \notag \\ 
    & = E \left [ \frac{1}{n}(Q_{\Sigma}(t+k(n-d))^2 - Q_{\Sigma}(t)^2) \right ] \notag \\
    & = \frac{1}{n} E \left [ \left ( Q_{\Sigma}(t) + \sum\limits_{j=0}^{k(n-d)-1} \sum\limits_{l=1}^n A_l(t+j) - \sum\limits_{j=0}^{k(n-d)-1} \sum\limits_{l=1}^n S_l(t+j) + \sum\limits_{j=0}^{k(n-d)-1} \sum\limits_{l=1}^n U_l(t+j) \right )^2 - Q_{\Sigma}(t)^2 \right ] \notag \\
    & = \frac{1}{n} \left \{ E \left [ \left ( Q_{\Sigma}(t) + \sum\limits_{j=0}^{k(n-d)-1} \sum\limits_{l=1}^n A_l(t+j) - \sum\limits_{j=0}^{k(n-d)-1} \sum\limits_{l=1}^n S_l(t+j) \right )^2 - Q_{\Sigma}(t)^2 \right ] \right. \notag \\
    & \quad + E \left [ 2\left ( Q_{\Sigma}(t) + \sum\limits_{j=0}^{k(n-d)-1} \sum\limits_{l=1}^n A_l(t+j) - \sum\limits_{j=0}^{k(n-d)-1} \sum\limits_{l=1}^n S_l(t+j) \right )\left ( \sum\limits_{j=0}^{k(n-d)-1} \sum\limits_{l=1}^n U_l(t+j) \right) \right ]  \notag \\
    & \quad \left. + E \left [ \left ( \sum\limits_{j=0}^{k(n-d)-1} \sum\limits_{l=1}^n U_l(t+j) \right)^2 \right ] \right \} \notag \\
    & = \frac{1}{n} \left \{ E \left [ 2 Q_{\Sigma}(t) \left (\sum\limits_{j=0}^{k(n-d)-1} \sum\limits_{l=1}^n A_l(t+j) - \sum\limits_{j=0}^{k(n-d)-1} \sum\limits_{l=1}^n S_l(t+j) \right ) \right ] \right. \notag \\
    & \quad + E \left [ \left (\sum\limits_{j=0}^{k(n-d)-1} \sum\limits_{l=1}^n A_l(t+j) - \sum\limits_{j=0}^{k(n-d)-1} \sum\limits_{l=1}^n S_l(t+j) \right ) ^2 \right ] \notag \\
    & \quad + E \left [ 2\left ( Q_{\Sigma}(t) + \sum\limits_{j=0}^{k(n-d)-1} \sum\limits_{l=1}^n A_l(t+j) - \sum\limits_{j=0}^{k(n-d)-1} \sum\limits_{l=1}^n S_l(t+j) \right )\left ( \sum\limits_{j=0}^{k(n-d)-1} \sum\limits_{l=1}^n U_l(t+j) \right) \right ]  \notag \\
    & \quad \left. + E \left [ \left ( \sum\limits_{j=0}^{k(n-d)-1} \sum\limits_{l=1}^n U_l(t+j) \right)^2 \right ] \right \} \notag
\end{align}
Since in steady-state we have $ E[\Delta W_{\parallel}(\tilde{\mathbf{Q}}(t')) ] = 0 $, then
\begin{align}
    & \frac{1}{n} E \left[
        Q_{\Sigma}(t) \left (\sum\limits_{j=0}^{k(n-d)-1} \sum\limits_{l=1}^n S_l(t+j) - \sum\limits_{j=0}^{k(n-d)-1} \sum\limits_{l=1}^n A_l(t+j) \right ) \right ] \notag \\
    & = \frac{1}{2n} \left \{ E \left [ \left (\sum\limits_{j=0}^{k(n-d)-1} \sum\limits_{l=1}^n A_l(t+j) - \sum\limits_{j=0}^{k(n-d)-1} \sum\limits_{l=1}^n S_l(t+j) \right ) ^2 \right ] \right. \notag \\
    & \quad + 2 E \left [ \left ( Q_{\Sigma}(t) + \sum\limits_{j=0}^{k(n-d)-1} \sum\limits_{l=1}^n A_l(t+j) - \sum\limits_{j=0}^{k(n-d)-1} \sum\limits_{l=1}^n S_l(t+j) \right )\left ( \sum\limits_{j=0}^{k(n-d)-1} \sum\limits_{l=1}^n U_l(t+j) \right) \right ]  \notag \\
    & \quad \left. + E \left [ \left ( \sum\limits_{j=0}^{k(n-d)-1} \sum\limits_{l=1}^n U_l(t+j) \right)^2 \right ] \right \} \label{eq:upb_equation}
\end{align}
We bound the second term in Equation \eqref{eq:upb_equation} as follows:
\begin{align}
    & \frac{1}{n} \left ( Q_{\Sigma}(t) + \sum\limits_{j=0}^{k(n-d)-1} \sum\limits_{l=1}^n A_l(t+j) - \sum\limits_{j=0}^{k(n-d)-1} \sum\limits_{l=1}^n S_l(t+j) \right )\left ( \sum\limits_{j=0}^{k(n-d)-1} \sum\limits_{l=1}^n U_l(t+j) \right) \label{eq:upb_qu_org} \\
    & = \frac{1}{n} \left ( Q_{\Sigma}(t) + \sum\limits_{j=0}^{k(n-d)-1} \sum\limits_{l=1}^n A_l(t+j) - \sum\limits_{j=0}^{k(n-d)-1} \sum\limits_{l=1}^n S_l(t+j) + \sum\limits_{j=0}^{k(n-d)-1} \sum\limits_{l=1}^n U_l(t+j) - \sum\limits_{j=0}^{k(n-d)-1} \sum\limits_{l=1}^n U_l(t+j) \right ) \notag \\
    & \quad \times \left ( \sum\limits_{j=0}^{k(n-d)-1} \sum\limits_{l=1}^n U_l(t+j) \right) \notag \\
    & \leq \frac{1}{n} \left ( Q_{\Sigma}(t) + \sum\limits_{j=0}^{k(n-d)-1} \sum\limits_{l=1}^n A_l(t+j) - \sum\limits_{j=0}^{k(n-d)-1} \sum\limits_{l=1}^n S_l(t+j) + \sum\limits_{j=0}^{k(n-d)-1} \sum\limits_{l=1}^n U_l(t+j) \right )\left ( \sum\limits_{j=0}^{k(n-d)-1} \sum\limits_{l=1}^n U_l(t+j) \right) \notag \\
    & = \frac{1}{n}  Q_{\Sigma}(t+k(n-d)) \left ( \sum\limits_{j=0}^{k(n-d)-1} \sum\limits_{l=1}^n U_l(t+j) \right) \notag \\
    & = \sum\limits_{l=1}^{n} \left [ Q_l(t+k(n-d)) \left ( \frac{\sum\limits_{j=0}^{k(n-d)-1} \sum\limits_{l=1}^n U_l(t+j)}{n} \right ) \right ] \notag \\
    & = \sum\limits_{l=1}^{n} \left [ Q_l(t+k(n-d)) \left ( \sum\limits_{j=0}^{k(n-d)-1} U_l(t+j) - \left ( \sum\limits_{j=0}^{k(n-d)-1} U_l(t+j) - \frac{\sum\limits_{j=0}^{k(n-d)-1} \sum\limits_{l=1}^n U_l(t+j)}{n} \right ) \right ) \right ] \notag \\
    & = \sum\limits_{l=1}^{n} \left [ Q_l(t+k(n-d)) \left ( \sum\limits_{j=0}^{k(n-d)-1} U_l(t+j) \right ) \right ] \notag \\
    & \quad - \sum\limits_{l=1}^{n} \left [ Q_l(t+k(n-d)) \left ( \sum\limits_{j=0}^{k(n-d)-1} U_l(t+j) - \frac{\sum\limits_{j=0}^{k(n-d)-1} \sum\limits_{l=1}^n U_l(t+j)}{n} \right ) \right ] \notag \\
    & \overset{(a)}{=} \sum\limits_{l=1}^{n} \left [ Q_l(t+k(n-d)) \left ( \sum\limits_{j=0}^{k(n-d)-1} U_l(t+j) \right ) \right ] \notag \\
    & \quad - \sum\limits_{l=1}^{n} \left [ \left (Q_l(t+k(n-d)) - \frac{Q_{\Sigma}(t+k(n-d))}{n} \right ) \left ( \sum\limits_{j=0}^{n-d-1} U_l(t+j) - \frac{\sum\limits_{j=0}^{k(n-d)-1} \sum\limits_{l=1}^n U_l(t+j)}{n} \right ) \right ] \notag \\
    &\overset{(b)}{=} \sum\limits_{l=1}^{n} \left [ Q_l(t+k(n-d)) \left ( \sum\limits_{j=0}^{k(n-d)-1} U_l(t+j) \right ) \right ] \notag \\ 
    & \quad - \sum\limits_{l=1}^{n} \left [ \left (Q_l(t+k(n-d)) - \frac{Q_{\Sigma}(t+k(n-d))}{n} \right ) \left ( \sum\limits_{j=0}^{k(n-d)-1} U_l(t+j) \right ) \right ] \notag \\
    & = \sum\limits_{l=1}^{n} \left [ Q_l(t+k(n-d)) \left ( \sum\limits_{j=0}^{k(n-d)-1} U_l(t+j) \right ) \right ] \notag \\
    & \quad + \sum\limits_{l=1}^{n} \left [ \left ( \frac{Q_{\Sigma}(t+k(n-d))}{n} - Q_l(t+k(n-d)) \right ) \left ( \sum\limits_{j=0}^{k(n-d)-1} U_l(t+j) \right ) \right ], \label{eq:upb_qu}
\end{align}
where (a) and (b) follow from adding and subtracting (respectively) the following zero-value terms
\begin{align}
    & \sum\limits_{l=1}^n \left [ \left ( \frac{Q_{\Sigma}(t+k(n-d))}{n} \right ) \left ( \sum_{j=0}^{k(n-d)-1}U_l(t+j) - \frac{\sum\limits_{j=0}^{k(n-d)-1} \sum\limits_{l=1}^n U_l(t+j)}{n} \right ) \right ] \notag \\
    & \qquad \qquad = \left ( \frac{Q_{\Sigma}(t+k(n-d))}{n} \right ) \sum\limits_{l=1}^n \left [ \left ( \sum_{j=0}^{k(n-d)-1}U_l(t+j) - \frac{\sum\limits_{j=0}^{k(n-d)-1} \sum\limits_{l=1}^n U_l(t+j)}{n} \right ) \right ] \notag \\
    & \qquad \qquad = \left ( \frac{Q_{\Sigma}(t+k(n-d))}{n} \right ) \left ( \sum\limits_{j=0}^{k(n-d)-1} \sum\limits_{l=1}^n U_l(t+j) - \sum_{j=0}^{k(n-d)-1} \sum_{l=1}^n U_l(t+j) \right ) \notag \\
    & \qquad \qquad = 0 \notag \\
    & \sum\limits_{l=1}^n \left [ \left ( Q_l(t+k(n-d)) - \frac{Q_{\Sigma}(t+k(n-d))}{n} \right ) \left ( \frac{\sum\limits_{j=0}^{k(n-d)-1} \sum\limits_{l=1}^n U_l(t+j)}{n} \right ) \right ] \notag \notag \\
    & \qquad \qquad = \left ( \frac{\sum\limits_{j=0}^{k(n-d)-1} \sum\limits_{l=1}^n U_l(t+j)}{n} \right ) \sum\limits_{l=1}^n \left [ \left ( Q_l(t+k(n-d)) - \frac{Q_{\Sigma}(t+k(n-d))}{n} \right ) \right ] \notag \\
    & \qquad \qquad = \left ( \frac{\sum\limits_{j=0}^{k(n-d)-1} \sum\limits_{l=1}^n U_l(t+j)}{n} \right ) \Big[ Q_{\Sigma}(t+k(n-d)) - Q_{\Sigma}(t+k(n-d)) \Big] \notag \\
    & \qquad \qquad = 0 \notag
\end{align}
For the first term in Equation \eqref{eq:upb_qu}, we bound
\[ Q_l(t+k(n-d)) \left( \sum_{j=0}^{k(n-d)-1} U_l(t+j) \right ) \]
for each server $l$. For each $ l $, we have
\[ Q_l(t+j+1)U_l(t+j) = 0 \]
and
\[ Q_l(t+j+1) = Q_l(t+j) + A_l(t+j) - S_l(t+j) + U_l(t+j), \]
for all $j \in \{0,1,...,k(n-d)-1\}$. It follows that
\begin{align}
    & Q_l(t+k(n-d)) \left( \sum\limits_{j=0}^{k(n-d)-1} U_l(t+j) \right ) \notag \\
    & =  Q_l(t+k(n-d)) \left( \sum\limits_{j=0}^{k(n-d)-2} U_l(t+j) \right ) + Q_l(t+k(n-d)) U_l(t+k(n-d)-1) \notag \\
    & = Q_l(t+k(n-d)-1) \left( \sum\limits_{j=0}^{k(n-d)-2} U_l(t+j) \right ) \notag \\
    & \quad + \Big[A_l(t+k(n-d)-1) - S_l(t+k(n-d)-1)+U_l(t+k(n-d)-1) \Big] \left( \sum\limits_{j=0}^{k(n-d)-2} U_l(t+j) \right ) \notag \\
    & = Q_l(t+1) U_l(t) + \sum\limits_{i=1}^{k(n-d)-1}\left [ (A_l(t+i) -S_l(t+i)+U_l(t+i)) \left( \sum\limits_{j=0}^{i-1} U_l(t+j) \right ) \right ] \notag \\
    & = \sum\limits_{i=1}^{k(n-d)-1}\left [ (A_l(t+i) -S_l(t+i) + U_l(t+i)) \left( \sum\limits_{j=0}^{i-1} U_l(t+j) \right ) \right ] \notag \\
    & \overset{(a)}{\leq} \sum\limits_{i=1}^{k(n-d)-1}\left [ A_l(t+i) \left( \sum\limits_{j=0}^{i-1} U_l(t+j) \right ) \right ] \notag \\
    & \leq \sum\limits_{i=1}^{k(n-d)-1}\left [ A_l(t+i) \left( \sum\limits_{j=0}^{k(n-d)-1}\sum\limits_{k=1}^{n} U_k(t+j) \right ) \right ] \notag \\
    & = \left [ \begin{array}{ll}
        \sum\limits_{i^{\prime}} A_l(t+i^{\prime}) \left( \sum\limits_{j=0}^{k(n-d)-1}\sum\limits_{k=1}^{n} U_k(t+j) \right ) & \forall \ l: l \in I \setminus I_d(t) \\
        0 & \forall \ l: l \in I_d(t)
    \end{array} \right ., \label{eq:upb_alul}
\end{align}
where $i^{\prime}$ are the time slots in which the $l$-th queue receives exogenous arrivals, and where (a) follows from the fact that $S_l(t+j) \geq U_l(t+j)$ for all $\l\in [n]$ and $j\geq 1$. For all $n$ servers, we have
\begin{align}
    \sum\limits_{l=1}^{n}\left [ \left ( Q_l(t+k(n-d)) \right) \left( \sum\limits_{j=0}^{k(n-d)-1} U_l(t+j) \right ) \right ] & \leq \sum\limits_{l\in I \setminus I_d }^{n} \left [ \sum\limits_{i^{\prime}} \left( A_l[t+i^{\prime}] \right) \left( \sum\limits_{j=0}^{k(n-d)-1}\sum\limits_{k=1}^{n} U_k(t+j) \right ) \right ] \notag \\
    & \overset{(a)}{\leq} \sum\limits_{l\in I \setminus I_d }^{n} \left [ kn A_{\max} \left( \sum\limits_{j=0}^{k(n-d)-1}\sum\limits_{k=1}^{n} U_k(t+j) \right ) \right ] \notag \\
    & = (n-d) k n A_{\max} \left( \sum\limits_{j=0}^{k(n-d)-1}\sum\limits_{k=1}^{n} U_k(t+j) \right ), \notag
\end{align}
Where (a) follows from the fact that the number of arrivals is bounded by $ n A_{\max} $. Combining this with equations \eqref{eq:upb_qu_org} and \eqref{eq:upb_qu} we get
\begin{align}
    & \frac{1}{n} E \left[
        \left ( Q_{\Sigma}(t) + \sum\limits_{j=0}^{k(n-d)-1} \sum\limits_{l=1}^n A_l(t+j) - \sum\limits_{j=0}^{k(n-d)-1} \sum\limits_{l=1}^n S_l(t+j) \right )\left ( \sum\limits_{j=0}^{k(n-d)-1} \sum\limits_{l=1}^n U_l(t+j) \right) \right ] \notag \\
    & \leq E \left[ \sum\limits_{l=1}^{n} \left [ \left ( Q_l(t+k(n-d)) \right )\left ( \sum\limits_{j=0}^{k(n-d)-1} U_l(t+j) \right ) \right ] \right ] \notag \\
    & \quad + E \left[ \sum\limits_{l=1}^{n} \left [ \left ( \frac{Q_{\Sigma}(t+k(n-d))}{n} - Q_l(t+k(n-d)) \right ) \left ( \sum\limits_{j=0}^{k(n-d)-1} U_l(t+j) \right ) \right ] \right ] \notag \\ 
    & \leq k(n-d)n A_{\max} E \left[ \left( \sum\limits_{j=0}^{k(n-d)-1}\sum\limits_{k=1}^{n} U_k(t+j) \right ) \right ] \notag \\
    & \quad + E \left[ \sum\limits_{l=1}^{n} \left [ \left ( \frac{Q_{\Sigma}(t+k(n-d))}{n} - Q_l(t+k(n-d)) \right ) \left ( \sum\limits_{j=0}^{k(n-d)-1} U_l(t+j) \right ) \right ] \right ] \notag \\
    & \overset{(a)}{\leq} k(n-d)n A_{\max} E \left[ \left( \sum\limits_{j=0}^{k(n-d)-1}\sum\limits_{k=1}^{n} U_k(t+j) \right ) \right ]  \notag \\
    & \quad + \sqrt{E \left[ \sum\limits_{l=1}^{n} \left ( \frac{Q_{\Sigma}(t+k(n-d))}{n} - Q_l(t+k(n-d)) \right)^2 \right ] E \left[ \sum\limits_{l=1}^{n} \left ( \sum\limits_{j=0}^{k(n-d)-1} U_l(t+j) \right )^2 \right ] }  \notag \\
    & \overset{(b)}{=} k(n-d)n A_{\max} 
        E \left[ \left( \sum\limits_{j=0}^{k(n-d)-1}\sum\limits_{k=1}^{n} U_k(t+j) \right ) \right ] 
        + \sqrt{E \left[ \sum\limits_{l=1}^{n} \left ( \frac{Q_{\Sigma}(t)}{n} - Q_l(t) \right)^2 \right ]
        E \left[ \sum\limits_{l=1}^{n} \left ( \sum\limits_{j=0}^{k(n-d)-1} U_l(t+j) \right )
        ^2 \right ]
        }  \notag \\ 
    & = k(n-d)n A_{\max} 
        E \left[ \left( \sum\limits_{j=0}^{k(n-d)-1}\sum\limits_{k=1}^{n} U_k(t+j) \right ) \right ] 
        + \sqrt{E \left[ \| Q_{\perp} \|_2^2 \right ]
        E \left[ \sum\limits_{l=1}^{n} \left ( \sum\limits_{j=0}^{k(n-d)-1} U_l(t+j) \right )
        ^2 \right ] },  \label{eq:upb_Eofqu}
\end{align}
where (a) follows from the Cauchy-Schwartz inequality, and (b) follows from the fact that $\tilde{\mathbf{Q}}(t'+1) \overset{d}{=} \tilde{\mathbf{Q}}(t') $ because they are in steady-state. To bound the first term in Equation \eqref{eq:upb_Eofqu}, we use that
\begin{align}
    0&= E \left [ \frac{\tilde Q_{\Sigma}(t'+1)}{\sqrt{n}} - \frac{\tilde Q_{\Sigma}(t')}{\sqrt{n}} \right ] \notag \\ 
    & = E \left [ \frac{Q_{\Sigma}(t+k(n-d))}{\sqrt{n}} - \frac{Q_{\Sigma}(t)}{\sqrt{n}} \right ]  \notag \\ 
    & = E \left [ \frac{ Q_{\Sigma}(t) + \sum\limits_{j=0}^{k(n-d)-1} \sum\limits_{l=1}^n A_l(t+j) - \sum\limits_{j=0}^{k(n-d)-1} \sum\limits_{l=1}^n S_l(t+j) + \sum\limits_{j=0}^{k(n-d)-1} \sum\limits_{l=1}^n U_l(t+j) }{\sqrt{n}} - \frac{Q_{\Sigma}(t)}{\sqrt{n}} \right ], \notag
\end{align}
and thus
\begin{align}
    E \left [ \sum\limits_{j=0}^{k(n-d)-1} \sum\limits_{l=1}^n U_l(t+j) \right ] &= E \left [ \sum\limits_{j=0}^{k(n-d)-1} \sum\limits_{l=1}^n S_l(t+j) - \sum\limits_{j=0}^{k(n-d)-1} \sum\limits_{l=1}^n A_l(t+j) \right ] \notag \\
    & = k(n-d)\epsilon. \label{eq:upb_Eofu}
\end{align}
To bound the second term in Equation \eqref{eq:upb_Eofqu}, we use that
\begin{align}
    E \left[ \sum\limits_{l=1}^{n} \left ( \sum\limits_{j=0}^{k(n-d)-1} U_l(t+j) \right )
        ^2 \right ] & \leq E \left[ \sum\limits_{l=1}^{n} \left [ \left ( \sum\limits_{j=0}^{k(n-d)-1} U_l(t+j) \right )
        k(n-d)S_{\max} \right ] \right ] \notag \\
    & = k(n-d)S_{\max} E \left [ \sum\limits_{j=0}^{k(n-d)-1} \sum\limits_{l=1}^n U_l(t+j) \right ]  \notag \\
    & = k^2(n-d)^2 S_{\max}\epsilon \label{eq:upb_Eofu2}
\end{align}

Using the bounds in equations \eqref{eq:upb_Eofqu}, \eqref{eq:upb_Eofu}, and \eqref{eq:upb_Eofu2}, we bound each term in Equation \eqref{eq:upb_equation} by
\begin{align}
    E \left[
        \left (\sum\limits_{j=0}^{k(n-d)-1} \sum\limits_{l=1}^n A_l(t+j) - \sum\limits_{j=0}^{k(n-d)-1} \sum\limits_{l=1}^n S_l(t+j) \right ) ^2 \right ] 
    \leq k^2(n-d)^2 \epsilon ^2 + k(n-d)n \sigma_{\lambda}^2 + k(n-d) \left ( \sum\limits_{l=1}^{n}\sigma_{\mu_l}^2 \right ) \notag
\end{align}
\begin{align}
    E \left[ \left ( \sum\limits_{j=0}^{k(n-d)-1} \sum\limits_{l=1}^n U_l(t+j) \right)^2 \right ] & \leq k^2(n-d)^2nS_{max}\epsilon \notag 
\end{align}
\begin{align}
    & \frac{1}{n} E \left[
        \left ( Q_{\Sigma}(t) + \sum\limits_{j=0}^{k(n-d)-1} \sum\limits_{l=1}^n A_l(t+j) - \sum\limits_{j=0}^{k(n-d)-1} \sum\limits_{l=1}^n S_l(t+j) \right )\left ( \sum\limits_{j=0}^{k(n-d)-1} \sum\limits_{l=1}^n U_l(t+j) \right) \right ] \notag \\
    & \leq k(n-d) n A_{\max} 
        E \left[ \sum\limits_{j=0}^{k(n-d)-1}\sum\limits_{k=1}^{n} U_k(t+j) \right ] 
        + \sqrt{E \left[ \| Q_{\perp} \|_2^2  \right ]
        E \left[ \sum\limits_{l=1}^{n} \left ( \sum\limits_{j=0}^{k(n-d)-1} U_l(t+j) \right )
        ^2 \right ]
        }  \notag \\
    & \overset{(a)}{=} k(n-d) n A_{\max}k(n-d)\epsilon + \sqrt{N_2(n,k,d) k^2(n-d)^2nS_{\max}\epsilon} \notag \\
    & = k^2(n-d)^2n A_{\max}\epsilon + k(n-d)\sqrt{N_2(n,k,d) n S_{\max}\epsilon} \notag
\end{align}
where (a) follows from Theorem \ref{thm_our:SSC}. Finally, substituting the three inequalities above into the equation \eqref{eq:upb_equation}, and using that
\begin{align}
    E \left[ Q_{\Sigma}(t) \left (\sum\limits_{j=0}^{k(n-d)-1} \sum\limits_{l=1}^n S_l(t+j) - \sum\limits_{j=0}^{k(n-d)-1} \sum\limits_{l=1}^n A_l(t+j) \right ) \right ] 
    = k(n-d)\epsilon E \left[ Q_{\Sigma}(t) \right ], \notag 
\end{align}
we get
\begin{align}
    \frac{k(n-d)\epsilon E \left[ Q_{\Sigma}(t) \right ]}{n}  & \leq \frac{k^2(n-d)^2\epsilon^2 + k(n-d)n \sigma_{\lambda}^2 + k(n-d)\left ( \sum\limits_{l=1}^{n}\sigma_{\mu_l}^2 \right )}{2n} \notag \\
    & \quad + k^2(n-d)^2n A_{\max}\epsilon + k(n-d)\sqrt{N_2(n,k,d) n S_{\max}\epsilon} \notag \\
    & \quad + \frac{k^2(n-d)^2 n S_{max}\epsilon}{2n}, \notag
\end{align}
and thus
\begin{align}
    \epsilon E \left[ Q_{\Sigma}(t) \right ] \notag \leq
    \frac{n \sigma_{\lambda}^2 + \sum\limits_{l=1}^{n}\sigma_{\mu_l}^2 }{2} + \epsilon^2 \frac{k(n-d)}{2} + \epsilon k(n-d)n \frac{2n A_{\max}+S_{\max}}{2} + \sqrt{\epsilon} n\sqrt{N_2(n,k,d) n S_{\max}}. \notag 
\end{align}

\end{document}